\documentstyle[epsfig,wrapfig]{article}
\newcommand{\be}{\begin{equation}}
\newcommand{\ee}{\end{equation}}
\newcommand{\ba}{\begin{eqnarray}}
\newcommand{\ea}{\end{eqnarray}}
\newcommand{\la}{\langle}
\newcommand{\ra}{\rangle}
\newcommand{\di}{{\rm d}}
\newcommand{\lsim}{\renewcommand{\arraystretch}{0.4}
                   \begin{array}{c} < \\ \sim \end{array}}
\newcommand{\h}{\hat{\bf {P}}_{h \perp}}
\begin{document}
%============= TITLE, AUTHORS, AFFILIATION, PRE-PRINT NUMBER =======
\title{	Azimuthal single spin asymmetries in SIDIS\\
	in the light of chiral symmetry	breaking}
\author{P.~Schweitzer$^{a,b}$, A.~Bacchetta$^c$\\
        \footnotesize\it $^a$ Dipartimento di Fisica Nucleare e Teorica,
	Universit\`a degli Studi di Pavia, Pavia, Italy\\
	\footnotesize\it $^b$ Institut f\"ur Theoretische Physik II,
 	Ruhr-Universit\"at Bochum,  D-44780 Bochum, Germany\\
	\footnotesize\it $^c$ Institut f\"ur Theoretische Physik,
	Universit\"at Regensburg, D-93040 Regensburg, Germany}
\date{September 2003}
\maketitle
%\vspace{-9cm}\begin{flushright} {\tt version 1.1} \end{flushright}
%\vspace{7.5cm}
%============= ABSTRACT ============================================
\begin{abstract}
\noindent
An attempt is made to understand the $z$-dependence of the azimuthal single
spin asymmetries observed by the HERMES collaboration in terms of chiral
models based on effective quark and Goldstone boson degrees of freedom.
The effects of respectively neglecting and considering Gaussian intrinsic
parton transverse momenta and the Sivers effect are explored. Predictions
for the transverse target polarization experiment at HERMES are presented.
\end{abstract}

%====== SECTION 1: INTRODUCTION ====================================
\section{Introduction}
\label{Sec-1:Introduction}

The HERMES
\cite{Avakian:1999rr,Airapetian:1999tv,Airapetian:2001eg,Airapetian:2002mf},
CLAS \cite{Avakian:2003pk} and SMC \cite{Bravar:1999rq} collaborations
reported the observation of  nonzero single spin azimuthal asymmetries (SSA)
in semi-inclusive deep-inelastic scattering (SIDIS).
Single spin asymmetries in hard processes cannot be explained by means of
perturbative QCD \cite{Kane:nd}, rather they signal the appearance of
nonperturbative effects described in terms of so far unexplored distribution
and fragmentation functions and effects associated with parton transverse
momenta and quark orbital angular momenta.

In a factorized picture \cite{Mulders:1995dh,Boer:1997nt}, the SSA in SIDIS
can be explained in terms of the chirally odd twist-2 and twist-3 distribution
functions $h_1$, $h_L$ and $e$ \cite{transversity}, which appear in connection
with the Collins fragmentation function $H_1^{\perp}$ 
\cite{Collins:1992kk,Efremov:1992pe}, 
and the chirally even Sivers distribution function $f_{1T}^{\perp}$ 
\cite{Sivers:1989cc,Collins:2002kn,Brodsky:2002cx,Belitsky:2002sm,Anselmino:1994tv},
$H_1^{\perp a}$ describes the left--right asymmetry in the fragmentation
of a transversely polarized quark into an unpolarized hadron (Collins effect),
while $f_{1T}^{\perp a}$ quantifies the distribution of unpolarized quarks in
a transversely polarized nucleon (Sivers effect).
Both are referred to as T-odd,
i.e.\ they would vanish by time reversal invariance in the absence of
final-state
interactions.

The HERMES data on SSA from a longitudinally polarized target
\cite{Avakian:1999rr,Airapetian:1999tv,Airapetian:2001eg,Airapetian:2002mf}
were studied in
Refs.~\cite{DeSanctis:2000fh,Anselmino:2000mb,Efremov:2000za,Ma:2002ns,Efremov:2001cz,Efremov:2001ia}
in terms of the Collins effect only.
In these  approaches, the Sivers effect was neglected.
Different models and assumptions were explored in these works in order to
describe the $x$-dependence of the HERMES data, however, only one model has
been employed so far, namely the Collins ansatz \cite{Collins:1992kk}, for
the description of the $z$-dependence of the HERMES data.

In this note we shall attempt to describe the $z$-dependence of the HERMES
data using a different model for the Collins fragmentation function based
on a chirally invariant approach suggested in Ref.~\cite{Bacchetta:2002tk}.
The required information on the involved distribution functions will be
taken from the chiral quark-soliton model \cite{h1-model} in which --
as well as in a large class of other chiral models -- the Sivers function
vanishes \cite{Pobylitsa:2002fr}.
The combination of the two models is justified 
in the sense 
that both models describe the dynamics of strong interactions at low energies 
in terms of effective chiral quark and Goldstone boson degrees of freedom. 
Thus both models are essentially based on chiral symmetry breaking, which 
is known to play an important role in non-perturbative QCD in general, and 
in the T-odd fragmentation process in particular \cite{Collins:1992kk}.

The note is organized as follows. In Sec.~\ref{Sec-2:AUL} we will review
the relevant details of the HERMES single-spin asymmetry measurement.
In Sec.~\ref{Sec-3:AUL-model} we will compare the results of our model to the
HERMES data from the longitudinal target polarization experiment, assuming
vanishing (\ref{Sec-3.1:neglect-pT}) and Gaussian (\ref{Sec-3.2:AUL-Gauss})
intrinsic parton transverse momenta in the target.
In Sec.~\ref{Sec-4:Sivers} we shall make a rough estimate of how big the
Sivers effect could be to still be compatible with the HERMES data within
our approach. Finally, in Sec.~\ref{Sec-5:AUT} we will make predictions
for the HERMES transverse target polarization experiment.
In Sec.~\ref{Sec-6:conclusions} we summarize our work and conclude.

%====== SECTION: HERMES EXPERIMENT ===============================
\section{
The HERMES measurement of the {\boldmath$A_{UL}^{\sin\phi}$} asymmetry}
\label{Sec-2:AUL}

%--- FIGURE 1: KINEMATICS -----------------------------------------
\begin{figure}[t!]
	\begin{center}
	\includegraphics[width=7.5cm]{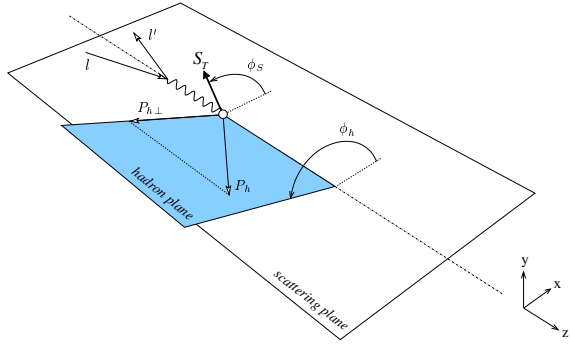}
	\end{center}
    	\caption{\footnotesize\sl Kinematics of
	the process $lp\rightarrow l'\pi X$ in the lab frame.}
	\end{figure}
%--- END FIGURE 1. ------------------------------------------------
In the HERMES experiments
\cite{Airapetian:1999tv,Airapetian:2001eg,Airapetian:2002mf} the cross section
for the process $lp\rightarrow l'h X$ was measured in dependence of the
azimuthal angle $\phi$ between lepton scattering plane and the production
plane of the hadron, see Fig.~1.

Let $P$, $l$ and $l'$ denote the momenta of target, incoming and
outgoing lepton, respectively. The kinematic variables --
center of mass energy $s$, four momentum transfer $q = l-l'$, invariant
mass of the photon-target system $W$, $x$, $y$ and $z$ -- are defined as
\begin{equation}
\begin{array}{rclrclrcl}
        s   &=& (P+l)^2 	\,, \qquad	&
        W^2  &=&  (P+q)^2 \,,\qquad	&
        Q^2  &=&  - q^2   	\,,	 \\
        x   &=& \frac{Q^2}{2Pq}  \,,\qquad	&
        y   & =&  \frac{Pq}{Pl}	\,,   \qquad  &
        z  &=&  \frac{PP_h}{Pq}
	\label{Eq:notation-1}		\;.
\end{array}
\end{equation}
In this notation the azimuthal asymmetry $A_{UL}^{\sin\phi}(z)$
studied by HERMES in the range $0.2 <  z  < 0.7$ reads
\be\label{Eq:AUL-00}
	A_{UL}^{\sin\phi}(z) =
  \frac{\displaystyle\int\!\!\di x\,\di y\,\di^2 {\bf P}_{h\perp}\;\sin\phi
	\left(
   	   \frac{1}{S^+}\,
   	   \frac{\di^5\sigma^+}{\di x\,\di y\,\di z\,\di^2 {\bf P}_{h\perp}}
	 - \frac{1}{S^-}\,
	   \frac{\di^5\sigma^-}{\di x\,\di y\,\di z\,\di^2 {\bf P}_{h\perp}}
	\right)}
        {\;\;\;\;\;\;\;\displaystyle
  	\frac{1}{2}\int\!\! \di x\,\di y\,\di^2 {\bf P}_{h\perp}\,
	\left(
   	   \frac{\di^5\sigma^+}{\di x\,\di y\,\di z\,\di^2 {\bf P}_{h\perp}}+
   	   \frac{\di^5\sigma^-}{\di x\,\di y\,\di z\,\di^2 {\bf P}_{h\perp}}
	\right)}\;\;.\ee
The subscript ``$U$'' reminds of the unpolarized beam, and
``$L$'' reminds of the longitudinally (with respect to the beam direction)
polarized proton target. $S^\pm$ denotes the modulus of target polarization
vector where ``$^+$'' means polarization opposite to the beam direction.
When integrating over $x$ and $y$ one has to consider the experimental cuts
\cite{Airapetian:1999tv,Airapetian:2001eg,Airapetian:2002mf}
\be\label{Eq:exp-cuts}
    	W > 2\,{\rm GeV}    	, \;\;\;
    	Q^2 > 1\,{\rm GeV}^2    , \;\;\;
	0.023 < x < 0.4	     \;	, \;\;\;
    	0.2   < y < 0.85     \; . \ee

The denominator in the asymmetry $A_{UL}^{\sin\phi}$ in Eq.(\ref{Eq:AUL-00})
is the cross section for pion production from scattering of an unpolarized
beam off an unpolarized target which -- after integrating out the transverse
momenta of the produced pions -- is given by
\be\label{Eq:expl-den}
    	\frac{1}{2} \int\!\!\di^2 {\bf P}_{h\perp}\left(
    	\frac{\di^5\sigma^+}{\di x\,\di y\,\di z\,\di^2{\bf P}_{h\perp}}+
    	\frac{\di^5\sigma^-}{\di x\,\di y\,\di z\,\di^2{\bf P}_{h\perp}}\right)
	\equiv \frac{\di^3\sigma_{UU}}{\di x\,\di y\,\di z}
    	= \frac{4\pi\alpha^2s}{Q^4}\;\left(1-y+\frac{y^2}{2}\right)\;
	  \sum_{a}e_a^2\,xf_1^a(x)\, D_1^a(z) \, .\ee

The cross sections entering the numerator in Eq.~(\ref{Eq:AUL-00})
were computed in Refs.~\cite{Mulders:1995dh,Boer:1997nt}
assuming that the process factorizes for ${\bf P}_{h\perp}^2 \ll Q^2$.
Arguments in favour of this assumption have been given \cite{Collins:1992kk},
however, a strict proof of a factorization theorem has not been presented yet.

The numerator in Eq.~(\ref{Eq:AUL-00}) consists of two parts -- originating
respectively from the longitudinal and the transverse component of the
target polarization vector with respect to the photon momentum ${\bf q}$
\be\label{Eq:expl-num}
	\int\!\!\di^2 {\bf P}_{h\perp}\,\sin\phi
	\left(
           \frac{1}{S^+}\,
	   \frac{\di^5\sigma^+}{\di x\,\di y\,\di z\,\di^2{\bf P}_{h\perp}}
    	 - \frac{1}{S^-}\,
	   \frac{\di^5\sigma^-}{\di x\,\di y\,\di z\,\di^2{\bf P}_{h\perp}}
    	\right)
	=\frac{2}{S}\;\frac{\di^3\sigma_{UL}}{\di x\,\di y\,\di z}
    	-\frac{2}{S}\;\frac{\di^3\sigma_{UT}}{\di x\,\di y\,\di z}
	\;. \ee
with
\ba
	\frac{\di^3\sigma_{UL}}{\di x\, \di y\,\di z}
	&=&
	- S_L \frac{4\pi\alpha^2 s}{Q^4}\;\frac{ M_N}{Q}\;2(2-y)\sqrt{1-y}\,
	\sum_a e_a^2\,x^2 \,{\cal I}\left[\frac{{\bf k}_T\cdot\h}{2 M_h}\,
	h_L^a(x,{\bf p}_T^2)\,H_1^{\perp a} (z,z^2 {\bf k}_T^2)\right]
	\nonumber \\
&&	\mbox{}
	\quad
	+ {\cal O}\Bigl(h_{1L}^\perp \widetilde{H}\Bigr)
	+ {\cal O}\Bigl(\frac{m_q}{M_N}\, g_1\, H_1^{\perp}\Bigr)\;,
	\label{Eq:sigmaUL-00}
	\\
	\frac{\di^4\sigma_{UT}}{\di x\, \di y\,\di z}
	&=&
	S_T \frac{4\pi\alpha^2 s}{Q^4}\;\sum_a e_a^2 \,x\,
	\Biggl\{-(1-y)\,{\cal I}\biggl[\frac{{\bf k}_T\cdot\h}{M_h}\,
	h_{1}^{a}(x,{\bf p}_T^2)H_1^{\perp a}(z,z^2 {\bf k}_T^2)\biggr]
	\nonumber \\
	&&\mbox{}
	\quad
	-(1-y+\frac{y^2}{2})\,{\cal I}\biggl[\frac{{\bf p}_T\cdot\h}{M_N}\,
	f_{1T}^{\perp a}(x,{\bf p}_T^2) D_1^a(z,z^2 {\bf k}_T^2) \biggr]
	\Biggr\}\;,
	\label{Eq:sigmaUT-00}
\ea
where $\h$ is a unit vector and we used the shorthand notation
\be
	{\cal I}\Bigl[\dots \Bigr] \equiv
   	\int \di^2 {\bf P}_{h \perp} \,\di^2  {\bf p}_T \,\di^2 {\bf k}_T \,
	\delta^{(2)}\Bigl({\bf p}_T
	-\frac{{\bf P}_{h \perp}}{z}-{\bf k}_T\Bigr) \;\Bigl[\dots \Bigr],
	\label{e:convolution}
\ee
The weight ``$\sin\phi$'' in Eq.~(\ref{Eq:AUL-00}) has the drawback of
leaving the unintegrated distribution and fragmentation functions inside a
convolution. The weight ``$\sin\phi\, |{\bf P}_{h\perp}|$'' would allow a
model-independent deconvolution \cite{Boer:1997nt}.

%====== SECTION 2: AUL =========================================
\section{Model calculation of the {\boldmath$A_{UL}^{\sin\phi}$} asymmetry}
\label{Sec-3:AUL-model}

In order to describe the HERMES data
we will use three ingredients:
Information on the Collins fragmentation function from the model calculation
of Ref.~\cite{Bacchetta:2002tk}.
Information on the involved (integrated) distribution functions from the
chiral quark-soliton model and the instanton vacuum model
\cite{h1-model,Dressler:1999hc}.
Models for the distribution of transverse quark momenta in the nucleon.

\paragraph{Collins fragmentation function \boldmath{$H_1^\perp$}.}
For the Collins fragmentation function we shall use the results
presented in Ref.~\cite{Bacchetta:2002tk}.
In that work, the Collins function has been estimated in a chiral invariant
approach \`a la Manohar and Georgi~\cite{Manohar:1984md},
where the effective degrees of freedom are
constituent quarks and pions, coupled via a pseudovector
interaction. In order to generate the
required phases, one-loop corrections to the quark propagator and vertex have
been included~\cite{Bacchetta:2002tk}.
In this approach, the unfavoured Collins function vanishes. (It would
appear only if one took into account two- and higher-loop corrections.)

In what follows we shall use the notation
\ba\label{Eq:favoured-frag}
	H_1^\perp
	&\equiv&
	  H_1^{\perp u/\pi^+} = H_1^{\perp\bar d/\pi^+}
	= H_1^{\perp d/\pi^-} = H_1^{\perp\bar u/\pi^-}
	= 2H_1^{\perp u/\pi^0} = 2H_1^{\perp d/\pi^0}
	= 2H_1^{\perp\bar u/\pi^0} = 2H_1^{\perp\bar d/\pi^0}\nonumber\\
	&\gg&
	  H_1^{\perp d/\pi^+} = H_1^{\perp\bar u/\pi^+}
	= H_1^{\perp u/\pi^-} = H_1^{\perp\bar d/\pi^-}
	\;.\ea
The first line of Eq.~(\ref{Eq:favoured-frag}) defines the favoured Collins
fragmentation function $H_1^\perp$ in terms of the fragmentation functions
for different flavours and pions charge conjugation and isospin symmetry
relations.  The second line of Eq.~(\ref{Eq:favoured-frag}) 
expresses the expectation that the unfavoured fragmentation is
suppressed with respect to the favoured fragmentation as it
has been conjectured on the basis of the Sch\"afer-Teryaev sum rule
\cite{Schafer:1999kn}. This conjecture remains, however, to be tested.

One should note that the assumption of favoured fragmentation cannot be
expected to work equally well for all pions. E.g., unfavoured fragmentation
effects have been shown to play an important role for $\pi^-$ production
from a proton target \cite{Ma:2002ns}, while in the case of $\pi^+$ production
$u$-quark dominance in the proton amplifies the effect of favoured
fragmentation.

Pioneering steps towards an understanding of the scale dependence
of the Collins function have been done in \cite{Henneman:2001ev}.
The results obtained there, however, need to be carefully reexamined in the
light of recent theoretical developments \cite{Goeke:2003az,Boer:2003cm}.
Therefore, for the ratio $H_1^\perp/D_1$ we shall use the result from
\cite{Bacchetta:2002tk}, which refers to a low scale below $1\,{\rm GeV}^2$
and assume that evolution to the average scale of the HERMES experiment can
be neglected.
While $D_1$ and possibly $H_1^\perp$ depend on the scale strongly,
one may hope that their ratio is less scale dependent.

\paragraph*{Chirally odd distribution functions
	   {\boldmath$h_1$} and {\boldmath$h_L$}.}
For the transversity distribution function $h_1^a$ we shall use predictions
\cite{h1-model} from the chiral quark-soliton model ($\chi$QSM).
The $\chi$QSM is a quantum field-theoretical relativistic model which was
derived from the instanton model of the QCD vacuum. The quark and antiquark
distribution functions obtained in this model satisfy all general QCD
requirements and agree, as far as they are known, to within (10-30)$\%$ with
phenomenological parameterizations \cite{Diakonov:1996sr}.
We shall assume that this model predicts $h_1^a(x)$ with a similar
uncertainty
\cite{h1-model}.

The twist-3 chirally odd distribution $h_L^a(x)$ is given by $h_L^a(x)=$
$2x\int_x^1\di y\,h_1^a(y)/y^2 + \widetilde{h}^a_L(x) +{\cal O}(m_q/ M_N)$.
In Ref. \cite{Dressler:1999hc} it was shown that in the instanton vacuum model
the (actual ``pure'') twist-3 term $\widetilde{h}^a_L(x)$ is strongly
suppressed with respect to the twist-2 part in the above-mentioned relation.
Thus, we can well approximate $h_L^a(x) = 2x\int_x^1\di y\,h_1^a(y)/y^2$ by
consistently neglecting $\widetilde{h}^a_L(x)$ and quark mass terms.

We will also need the deuteron transversity distribution
function which we shall estimate, e.g.\  for the $u$-quark,
as $h_1^{u/D}\approx h_1^{u/p}+h_1^{u/n}=h_1^u+h_1^d$, where isospin
symmetry was used in the last step (with $h_1^u\equiv h_1^{u/p}$, etc.).
Corrections due to the D-state admixture \cite{Umnikov:1996hy} are smaller
than other theoretical uncertainties in our approach and we disregard them
here.

The results for the chirally odd distribution functions are LO-evolved from
the low scale of the model to the average scale of the HERMES experiment of
$2.5\,{\rm GeV}^2$.
For the unpolarized distribution functions $f_1^a(x)$ we shall use
the parameterization of Ref.~\cite{GRV} at the corresponding scale.

In the $\chi$QSM -- as well as in a large class of other chiral soliton
models -- T-odd distribution functions vanish \cite{Pobylitsa:2002fr}.
Therefore in our approach it is consistent to neglect the Sivers function,
cf.\  below Section~\ref{Sec-3.2:AUL-Gauss}.

\paragraph{Transverse momentum distributions.}
While the transverse momentum distribution of the fragmenting quarks
is known from the model calculation of Ref.~\cite{Bacchetta:2002tk},
we have no information about ``unintegrated'' transverse momentum dependent
distribution functions\footnote{
	For a careful discussion of the precise meaning of ``unintegrated''
	transverse momentum dependent distribution functions in QCD
	see Ref.~\cite{Collins:2003fm}.}
from the $\chi$QSM. Such information, however, is required to describe the
HERMES data, as it is evident from Eqs.~(\ref{Eq:sigmaUL-00}) and
(\ref{Eq:sigmaUT-00}). In the following we shall use two different models,
namely
\begin{itemize}
\item{the neglect of intrinsic parton transverse momenta in the target, and}
\item{the assumption that intrinsic parton transverse momenta follow
      a Gaussian distribution.}
\end{itemize}

\subsection{Neglect of intrinsic {\boldmath $p_T$} in distribution functions}
\label{Sec-3.1:neglect-pT}

In this Section, we shall use a simple and extreme model.
Let $f(x,{\bf p}_T^2)$
be a generic distribution function, then we assume
\be\label{Eq:model-pT}
	f(x,{\bf p}_T^2) = f(x)\;\delta^{(2)}({\bf p}_T)\;.
\ee
This ansatz amounts essentially to the disregard of intrinsic quark transverse
momenta, abscribing the transverse momentum of the outgoing hadron entirely
to the fragmentation process.
The ansatz (\ref{Eq:model-pT}) immediately ``kills'' the Sivers effect,
however, it is not a too restrictive assumption in our approach, where
the Sivers distribution is zero anyway.
The cross sections of Eqs.~(\ref{Eq:sigmaUL-00}) and (\ref{Eq:sigmaUT-00})
become
\ba
	\frac{\di^3\sigma_{UL}}{\di x\, \di y\,\di z}
	&=&
	S_L \frac{4\pi\alpha^2 s}{Q^4}\;\frac{ M_N}{Q}\;2(2-y)\sqrt{1-y}\,
	    \sum_a e_a^2\,x^2 h_L^a(x)\,H_1^{\perp (1/2)a}(z)
            + {\cal O}\Bigl(h_{1L}^\perp \widetilde{H}\Bigr)
	    + {\cal O}\Bigl(\frac{m_q}{M_N}\Bigr)\;, \;\;\;\;
	\label{Eq:sigmaUL-01}
	\\
	\frac{\di^4\sigma_{UT}}{\di x\, \di y\,\di z}
	&=&
	S_T \frac{4\pi\alpha^2 s}{Q^4}\;(1-y)
    	    \sum_a e_a^2 x\,h_1^a(x)\,H_1^{\perp (1/2)a}(z) \;.
	\label{Eq:sigmaUT-01}
\ea
$S_L = S\,\cos\Theta_S$ and $S_T = S\,\sin\Theta_S$ are respectively the
longitudinal and transverse component of the target polarization $S$
with respect to the 3-momentum of the virtual photon, and
$\cos\Theta_S\simeq 1-2 M_N^2x(1-y)/(s y)$.
The transverse moment of the Collins fragmentation function in
Eqs.~(\ref{Eq:sigmaUL-01},~\ref{Eq:sigmaUT-01}) is defined as
\be
	H_1^{\perp (1/2)a}(z) = z^2 \int\!\!\di^2{\bf k_T}\;
	\frac{|{\bf k_T}|}{2 m_\pi}\;H_1^{\perp a}(z,z^2{\bf k}_{\bf T}^2)\;.
\ee
The term $\propto h_{1L}^\perp\widetilde{H}_1^\perp$ neglected in 
Eq.~(\ref{Eq:sigmaUL-01}) was estimated \cite{Efremov:2001cz} to be
small compared to the first term in Eq.~(\ref{Eq:sigmaUL-01}) in the 
kinematics of the HERMES experiment.
We also neglect quark mass effects.

Using charge conjugation and isospin symmetry and neglecting unfavoured
fragmentation we obtain for the azimuthal asymmetries
\be\label{Eq:AUL-z-final}
	A_{UL}^{\sin\phi}(z,\,\pi,\,{\rm target}) =
	C_L(\pi,\,\,{\rm target}) \; a^{(1/2)}(z) \;.
\ee
As a consequence of the favoured flavour fragmentation and the simplified
treatment of the deuteron target the $z$-dependence of the azimuthal
asymmetries for different pions from different targets is given by a
``universal'' function
\be\label{Eq:universal-a^(1/2)(z)}
	a^{(1/2)}(z) \equiv \frac{H_1^{\perp (1/2)}(z)}{D_1(z)}
	\;, \ee
while the information on the respective pion produced from the respective
target is contained in the constant
\ba
&&	\!\!\!\!\!\!
	C_L(\pi,\,{\rm target}) = 2\,\nonumber\\
&&	\times
	\frac{\int\!\di x\,\di y \sum_a^\pi e_a^2\,x/Q^4\,[
	    2\cos\Theta_S\,(2-y)\sqrt{1-y}( M_N/Q)\,x\,h_L^{a/{\rm target}}(x)
	    -\sin\Theta_S\,(1-y)                 \,x\,h_1^{a/{\rm target}}(x)]}
	     {\int\!\di x\,\di y\,(1-y+y^2/2)/Q^4
	      \sum_a^\pi e_a^2\,xf_1^{a/{\rm target}}(x)}
	\;\;\;\nonumber\\ \label{Eq:CL-pi-target}
\ea
where $\sum_a^\pi$ denotes the summation over the favoured flavours relevant
for the respective pion. Furthermore, the constants $C_L(\pi,\,{\rm target})$
depend on the experimental cuts which enter the integrations over $x$ and $y$
in (\ref{Eq:CL-pi-target}). The results for the constants
$C_L(\pi,\,\,{\rm target})$ are given in Table~1.
%
% BEGIN: TABLE
%
{\footnotesize\begin{table}
\begin{center}
	\begin{tabular}{|l||c|c||c|c|}
	\hline
  	&&&& \\
  	pion $\pi$ & $C_L(\pi,\,{\rm p})$ & $C_L(\pi,\,{\rm d})$
        	   & $C_T(\pi,\,{\rm p})$ & $C_T(\pi,\,{\rm d})$\\
  	&&&& \\
	\hline\hline&&&&\\
	$\pi^+$	 &  0.154 & 0.075 &  0.781 & 0.354 \\
	$\pi^0$	 &  0.109 & 0.063 &  0.536 & 0.297 \\
	$\pi^-$	 & -0.025 & 0.038 & -0.204 & 0.181 \\
	&&&&
	\\
\hline
\end{tabular}
\end{center}
	\caption{\footnotesize\sl
	The constants $C_L({\rm pion},\,{\rm target})$ and
	$C_T({\rm pion},\,{\rm target})$ as defined in
	Eqs.~(\ref{Eq:AUL-z-final},~\ref{Eq:CL-pi-target})
	and  (\ref{Eq:AUT-z},~\ref{Eq:CT-pi-target}), respectively.}
\end{table}}
%
% END: TABLE.
%

The results for the azimuthal asymmetries $A_{UL}^{\sin\phi}(z)$ are
shown as dotted lines in Figs.~2a-2f and compared to the HERMES data
\cite{Airapetian:1999tv,Airapetian:2001eg,Airapetian:2002mf}.
We conclude that, under the assumption (\ref{Eq:model-pT}) of vanishing
intrinsic quark transverse momenta, the Collins effect computed in
a chiral invariant approach can explain the data within the --
at present admittedly sizeable -- statistical error of the experiment.

%--- FIGURE 2: AUL(z) proton -----------------------------------------
	\begin{figure}[t!]
\begin{tabular}{ccc}
	\includegraphics[width=5.2cm,height=5cm]{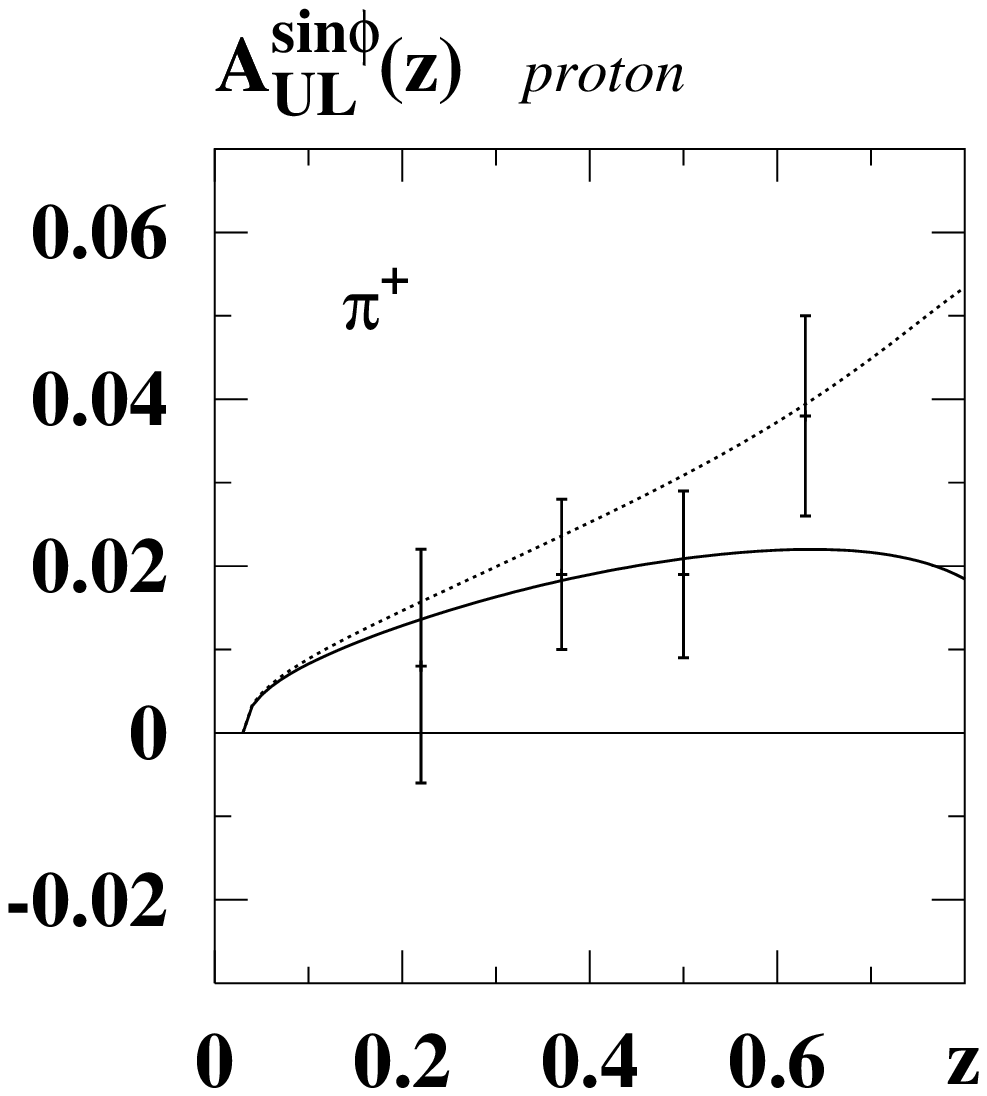} &
	\includegraphics[width=5.2cm,height=5cm]{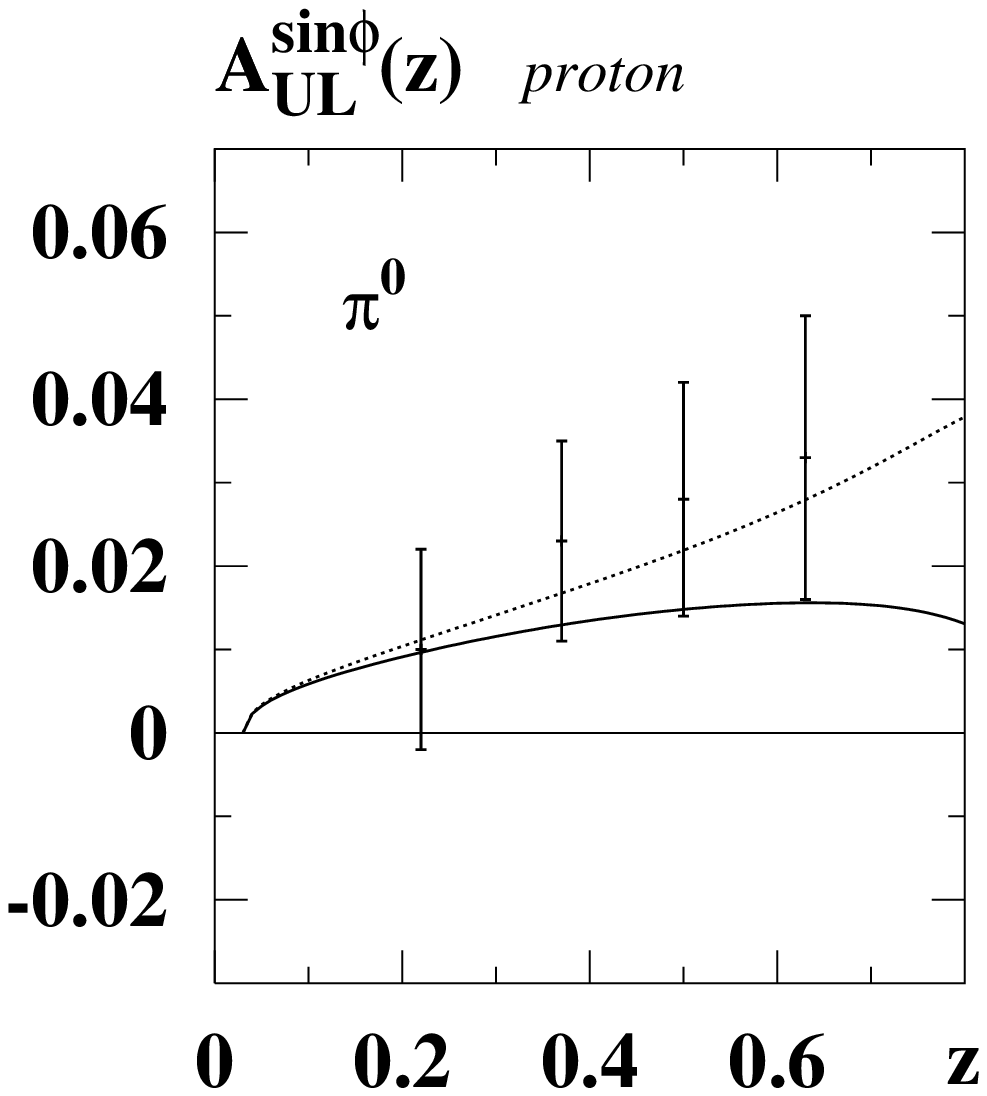} &
	\includegraphics[width=5.2cm,height=5cm]{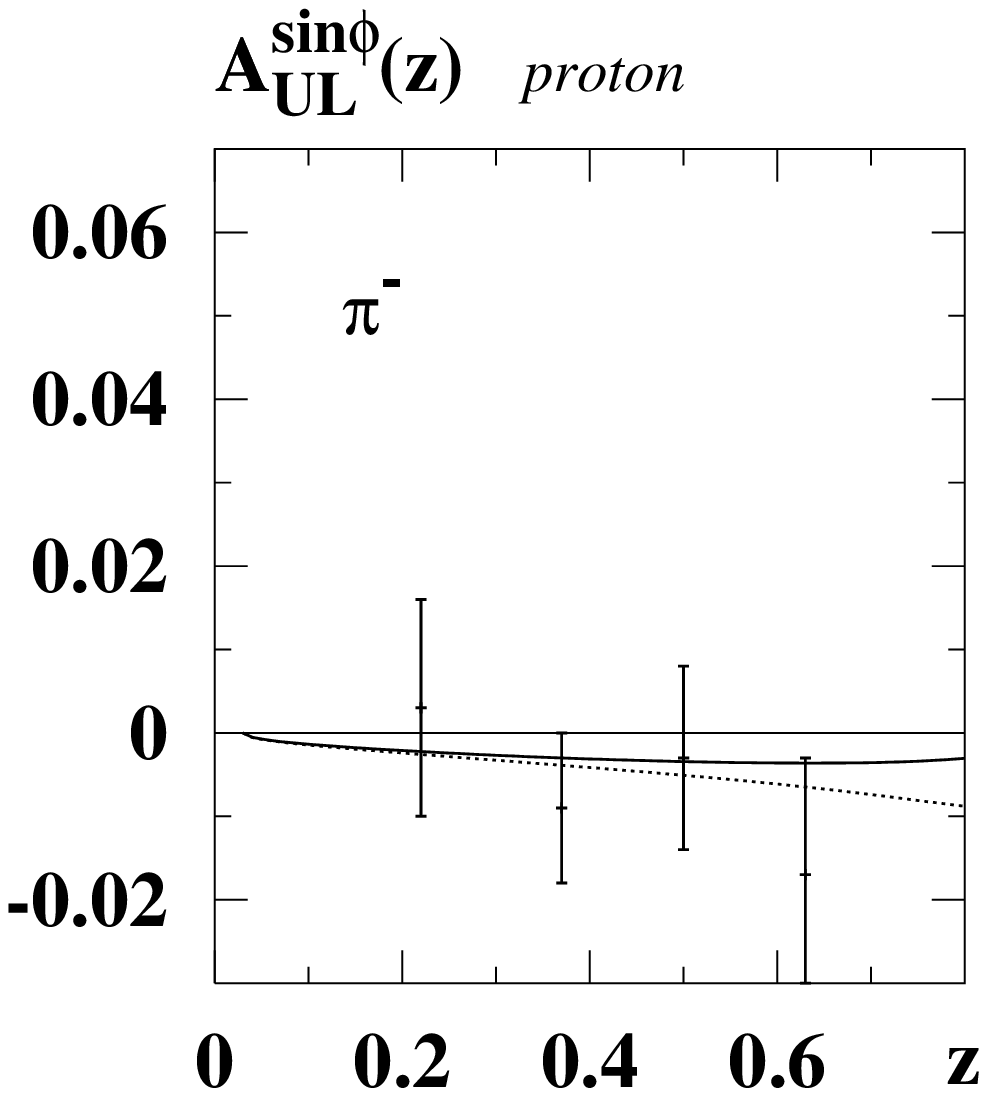}
	\cr
	{\bf a} &
	{\bf b} &
	{\bf c}
	\cr
	\includegraphics[width=5.2cm,height=5cm]{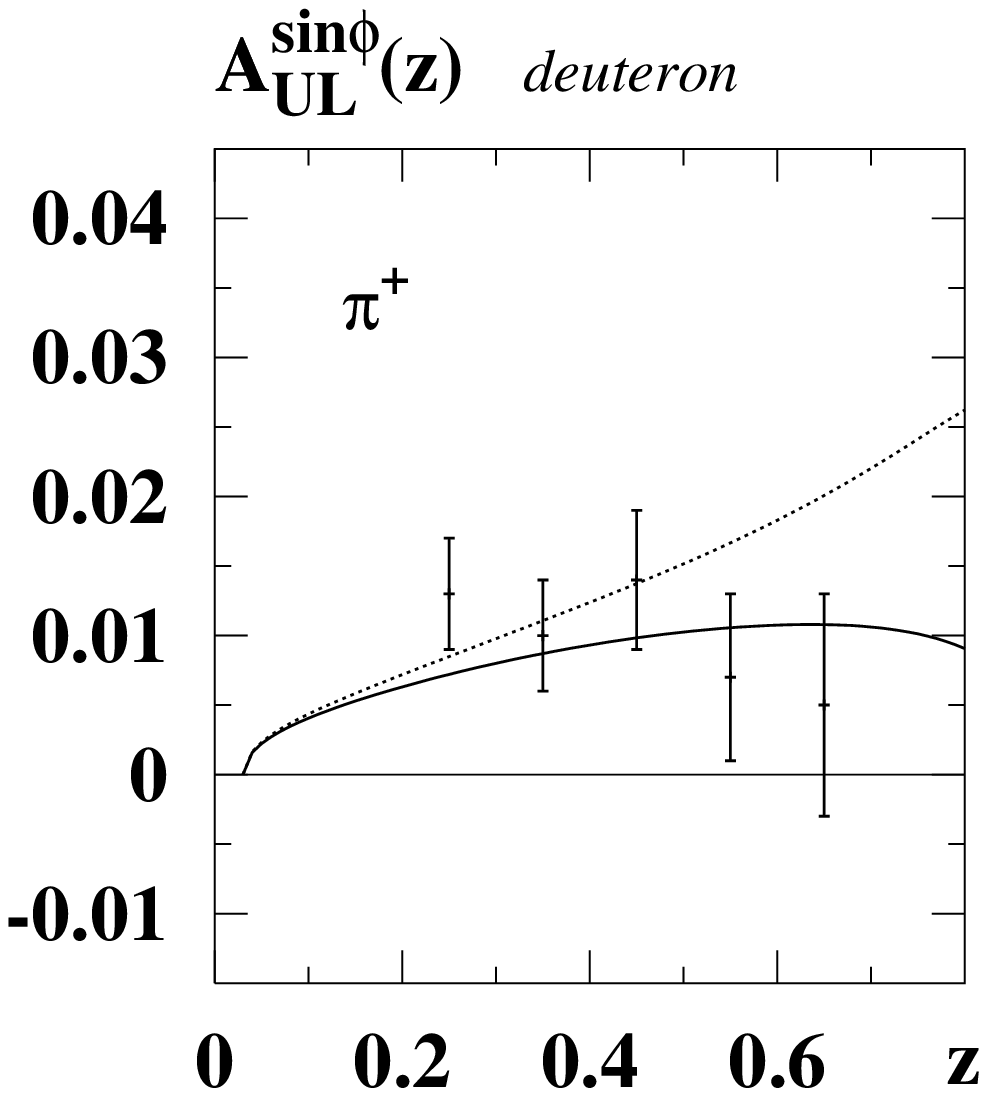} &
	\includegraphics[width=5.2cm,height=5cm]{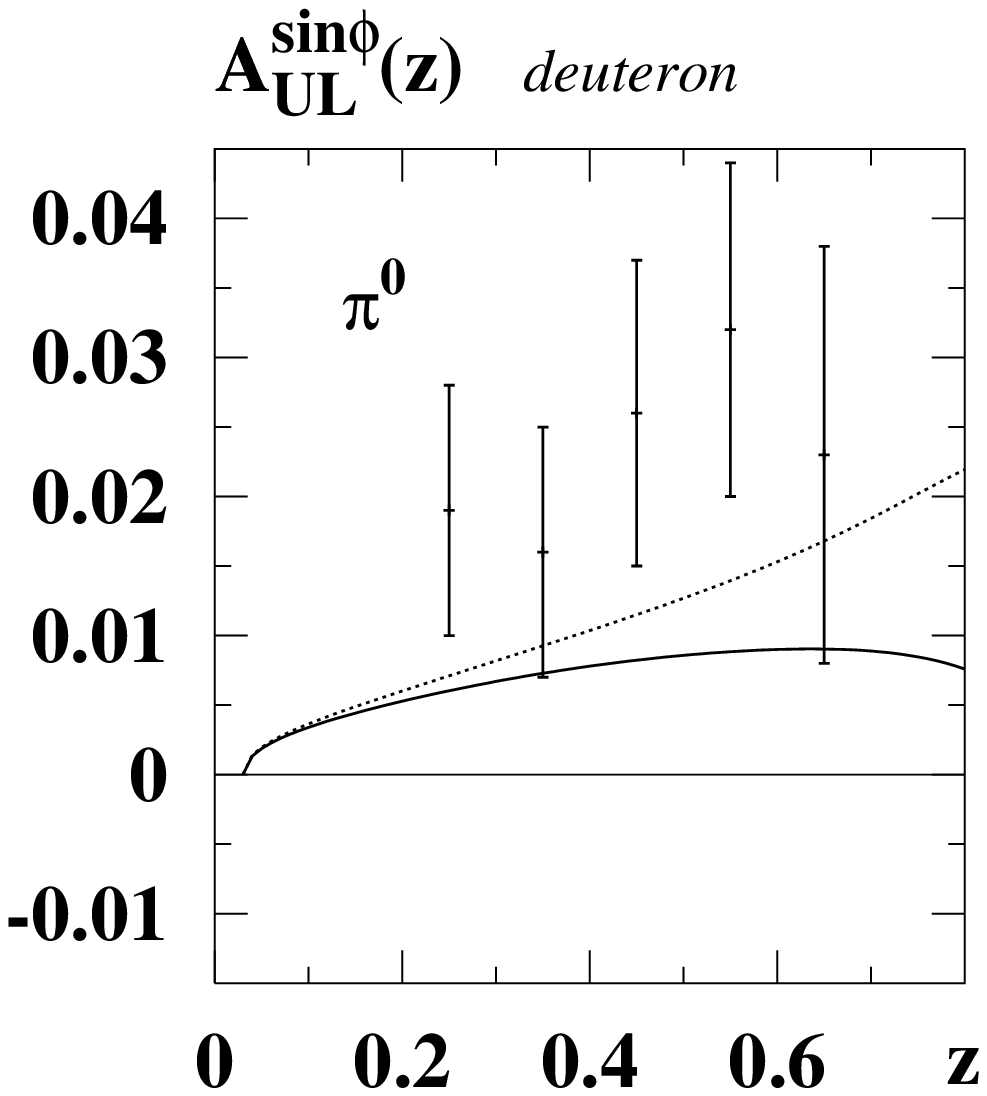} &
	\includegraphics[width=5.2cm,height=5cm]{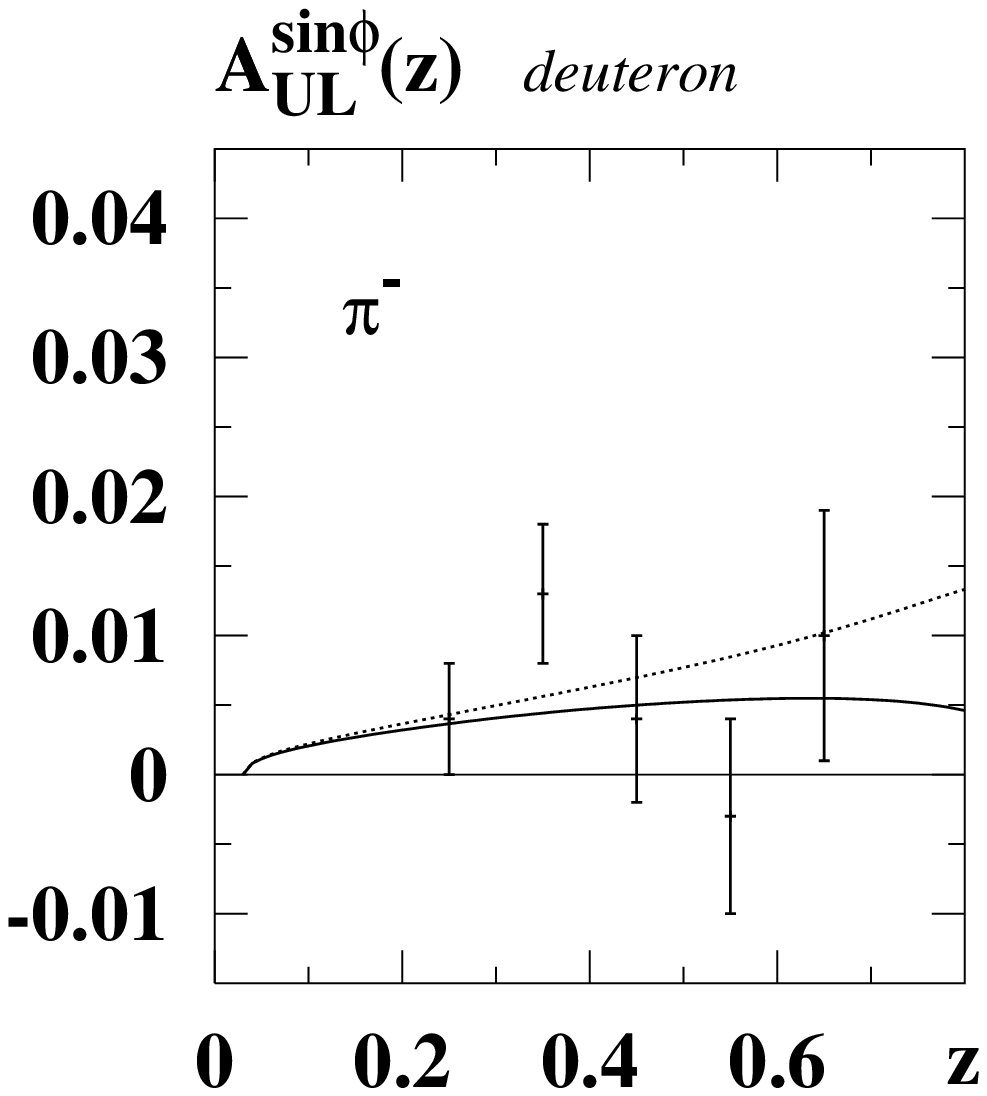}
	\cr
	{\bf d} &
	{\bf e} &
	{\bf f}
\end{tabular}
	\label{fig2-AUL}
	\caption{\footnotesize\sl
	The azimuthal single spin asymmetries $A_{UL}^{\sin\phi}(z)$ from
	a proton and deuteron target for $\pi^+$, $\pi^0$ and $\pi^-$ in
	comparison to the HERMES data
	\cite{Airapetian:1999tv,Airapetian:2001eg}.}
	\end{figure}
%--- END FIGURE 2: AUL(z)----- -----------------------------------------

%====== SECTION 4: Gaussian ansatz  ====================================
\subsection{Gaussian model for transverse quark momenta}
\label{Sec-3.2:AUL-Gauss}

The results presented in the previous Section rely on the assumption that
the intrinsic transverse momentum of partons in the target is zero.
In this section we try to estimate what is the effect of introducing a
nonzero intrinsic transverse momentum.
In order to do this, we will assume a Gaussian distribution of transverse
momentum, both for the distribution and fragmentation functions. Such an
assumption is in fair agreement with the HERMES data \cite{Airapetian:1999tv}.

Let $f(x,{\bf p}_T^2)$ and $D(z,z^2 {\bf k}_T^2)$ denote respectively
a generic unintegrated distribution and fragmentation function.
We assume that
\be\label{Eq:Gaussian-kT}
	f(x,{\bf p}_T^2) = \frac{f(x)}{\pi\la{\bf p}_T^2\ra}\;
	\exp\biggl(-\,\frac{{\bf p}_T^2}{\la{\bf p}_T^2\ra}\biggr)\;,\;\;\;
	D(z,{\bf K}_T^2) = \frac{D(z)}{\pi z^2 \la{\bf K}_T^2\ra}\;
	\exp\biggl(-\,\frac{{\bf K}_T^2}{\la{\bf K}_T^2\ra}\biggr)
	\ee
holds, where ${\bf K}_T= -z {\bf k}_T$ is the transverse momentum
the hadron acquires in the fragmentation process in the frame
where the fragmenting quark has no transverse momentum.
The functions are normalized such that
$\int\di^2{\bf p}_T f(x,{\bf p}_T^2) = f(x)$ and
$\int\di^2{\bf K}_T D(z,{\bf K}_T^2) = D(z)$.

Let us remark that the assumption (\ref{Eq:Gaussian-kT}) is not consistent
with the model result of Ref.~\cite{Bacchetta:2002tk}, which yields a
different transverse momentum distribution,
nor with the positivity bounds of Ref.~\cite{Bacchetta:1999kz}.
However, we do not address here the issue of the transverse momentum
dependence of the asymmetries but merely their $z$-dependence, and here
only averages over transverse momenta such as $H_1^{\perp (1/2)}(z)$ or
$\la{\bf K}_T^2(z)\ra$ are of relevance. At the level of such averages,
Eq.~(\ref{Eq:Gaussian-kT}) is compatible with the results from 
\cite{Bacchetta:2002tk} and can comply with the integrated 
positivity bounds, which is sufficient for our purposes.

The ansatz (\ref{Eq:Gaussian-kT}) is a convenient choice which allows
a disentanglement of the distribution and fragmentation functions in the
observed asymmetry. In fact, under the assumption (\ref{Eq:Gaussian-kT})
the azimuthal asymmetry is given by
\ba\label{AUL-Gaussian}
	\frac{\di^3\sigma_{UL}}{\di x\, \di y\,\di z}
	&=&
	S_L \frac{4\pi\alpha^2 s}{Q^4}\;\frac{ M_N}{Q}\;2(2-y)\sqrt{1-y\,}
	    \sum_a e_a^2\frac{x^2 h_L^a(x)\,H_1^{\perp (1/2)a}(z)}
	    {\sqrt{1+z^2\, \la{\bf p}_T^2(x)\ra/\la{\bf K}_T^2(z)\ra}} \;,
	\label{Eq:sigmaUL-02}
	\\
	\frac{\di^4\sigma_{UT}}{\di x\, \di y\,\di z}
	&=&
	S_T \frac{4\pi\alpha^2 s}{Q^4}\sum_a e_a^2
    	    \Biggl[(1-y)\,\frac{x\,h_1^a(x)\,H_1^{\perp (1/2)a}(z)}
	    {\sqrt{1+z^2\, \la{\bf p}_T^2(x)\ra/\la{\bf K}_T^2(z)\ra}}
	    \nonumber \\
	    &&\;\;\;\;\;\;\;\;\;\;\;\;
	    -(1-y+y^2/2)\,\frac{x\,f_{1T}^{\perp (1/2)a}(x)\,D_1^{a}(z)}
	    {\sqrt{1+\la{\bf K}_T^2(z)\ra/(z^2 \la{\bf p}_T^2(x)\ra})}\Biggr]
	\label{Eq:sigmaUT-02}\;.
\ea
The Sivers function appears in Eq.~(\ref{Eq:sigmaUT-02}) because the partons
in the target are now allowed to have non-vanishing intrinsic transverse
momenta. In our approach the Sivers function vanishes, however, in the next
Section we shall make use  of the explicit expressions with the Sivers effect
in Eq.~(\ref{Eq:sigmaUT-02}).
If we neglected transverse quark momenta in the target, i.e.\  if we took the
limit $\la p_T^2\ra\to 0$, we would recover the results of the previous
Section.

An important point is to reproduce the behaviour of 
$\la|{\bf P}_{h\perp}|(z)\ra$ observed in the HERMES experiment 
\cite{Airapetian:2002mf}.
If there were no transverse quark momenta in the target then
$\la |{\bf P}_{h\perp}|(z)\ra$ would arise from $\la|{\bf K}_T|(z)\ra$ only,
i.e. one would have $\la |{\bf P}_{h\perp}|(z)\ra = \la|{\bf K}_T|(z)\ra$.
In Fig.~3 we see that the $\la|{\bf K}_T|(z)\ra$ from \cite{Bacchetta:2002tk}
alone underestimates the HERMES data on $\la |{\bf P}_{h\perp}|(z)\ra$
\cite{Airapetian:2002mf} by about $30\%$.

This discrepancy could, of course, be attributed to the theoretical
uncertainty of the model calculation of Ref.~\cite{Bacchetta:2002tk},
in particular because the results of \cite{Bacchetta:2002tk} refer to
a low scale below $1\,{\rm GeV}^2$ while the HERMES data refer to
$\la Q^2\ra=2.5\,{\rm GeV}^2$. We have implicitly
taken such a point of view
in the previous Section, where transverse quark momenta in the
target were manifestly neglected.

Here she shall take an opposite point of view and assume that the
fragmentation transverse momentum is correctly described by the model
of Ref.~\cite{Bacchetta:2002tk} and determine the $\la{\bf p}_T^2\ra$
required to achieve a better description of $\la |{\bf P}_{h\perp}|(z)\ra$
at HERMES \cite{Airapetian:2002mf}.
The relation between ${\bf P}_{h\perp}$ of the hadron,
the intrinsic parton transverse momentum ${\bf p}_T$ in the target
and the transverse momentum ${\bf K}_T$ the hadron acquires in the
fragmentation process is given by \cite{Bacchetta:2002tk}
\be\label{Eq:Phperp-in-SIDIS}
	\la {\bf P}_{h\perp}^2(z)\ra
	= z^2\la{\bf p}_T^2\ra + \la{\bf K}_T^2(z)\ra
	\;.\ee
It is by no means clear how to use the relation (\ref{Eq:Phperp-in-SIDIS})
in order to describe $\la |{\bf P}_{h\perp}|(z)\ra$. If the distribution
of the transverse momenta of the produced hadrons were Gaussian, then
$\la {\bf P}_{h\perp}^2\ra=\frac{\pi}{4}\la |{\bf P}_{h\perp}|\ra^2$ and
\be\label{Eq:Phperp-in-SIDIS-2}
	\la |{\bf P}_{h\perp}|(z)\ra = \la|{\bf K}_T|(z)\ra\,
	\sqrt{1+z^2\la{\bf p}_T^2\ra / \la{\bf K}_T^2(z)\ra} \;.
	\ee
If we assume the relation (\ref{Eq:Phperp-in-SIDIS-2}) then we find that
$\la{\bf p}_T^2\ra=0.5\,{\rm GeV^2}$ allows to better describe the HERMES
data, see Fig.~3.
Two remarks are in order. Firstly, in general $\la{\bf p}_T^2\ra$
could be a function of $x$ which we disregard here for simplicity.
Secondly, a somehow smaller value of, say, $\la{\bf p}_T^2\ra=0.4\,{\rm GeV^2}$
would also allow to describe reasonably the data in Fig.~3.
However, we here we prefer to choose the larger value as an opposite to
the limiting case $\la{\bf p}_T^2\ra\to 0$ considered in the previous
Section~\ref{Sec-3.1:neglect-pT}.

Our estimate of $\la{\bf p}_T^2\ra=0.5\,{\rm GeV^2}$ lies in the range of the
values considered in literature: It somehow overestimates the results from
Refs.~\cite{Jackson:1989ph,Martin:1998sq,Abreu:1996na} and underestimates
the numbers $\la{\bf p}_T^2\ra \sim 0.8\,{\rm GeV}^2$ reported in
Ref.~\cite{D'Alesio:2003dv}.\footnote{
	It is worthwhile mentioning that this is the only place where we
	use an absolute number from the model of \cite{Bacchetta:2002tk}.
	In all other quantities ratios of model results enter, where
	theoretical uncertainties can, of course, add up but also have
	a chance to cancel.}

%--- FIGURE 3: Phperp(z) ------------------------------------------
\begin{figure}%[t!]
	\includegraphics[width=6.5cm,height=6.5cm]{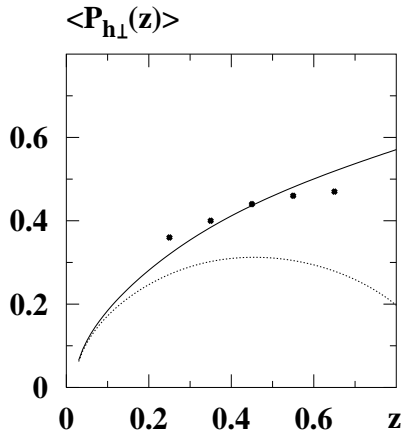}
    	\caption{\footnotesize\sl
  	The average transverse momentum of the produced hadrons
	$\la|{\bf P}_{h\perp}|(z)\ra$ as function of $z$.
	The dots are the $\la|{\bf P}_{h\perp}|(z)\ra$ HERMES data
	($h=$ pion) from Ref.~\cite{Airapetian:2002mf}. The curves follow
	from the model calculation of Ref.~\cite{Bacchetta:2002tk}
	if one assumes the parton transverse momenta in the target
	to be Gaussian (solid line) or to vanish (dashed line).}
	\end{figure}
%--- END FIGURE 3. ------------------------------------------------

The azimuthal asymmetry is given now by Eq.~(\ref{Eq:AUL-z-final}) with
$a^{(1/2)}(z)$ replaced by
\be\label{App:Gaussian-universal-a(z)}
	a_{\rm Gauss}^{(1/2)}(z) =
	\frac{1}{\sqrt{1+z^2 \la{\bf p}_T^2\ra/\la{\bf K}_T^2(z)\ra}\;}
	\frac{H_1^{\perp (1/2)}(z)}{D_1(z)} \;. \ee
Using $\la{\bf p}_T^2\ra=0.5\,{\rm GeV^2}$ we obtain the results for
$A_{UL}^{\sin\phi}(z)$ plotted as solid lines in Figs.~2a-2f. The description
of the HERMES data \cite{Airapetian:1999tv,Airapetian:2001eg,Airapetian:2002mf}
can be viewed as equally satisfactory as in the case discussed in
Section~\ref{Sec-3.1:neglect-pT}.

We conclude that the two different approaches -- the assumptions that
the distribution of parton transverse momenta in the target is negligible
and that it is Gaussian with a sizeable width -- cannot be discriminated
experimentally at present. One may expect that an optimized
phenomenological description would require a $\la{\bf p}_T^2\ra$ somewhere
between the limiting case $\la{\bf p}_T^2\ra\to 0$ and
$\la{\bf p}_T^2\ra=0.5\,{\rm GeV^2}$.

%====== SECTION 5: Sivers effect  ====================================
\section{Is there room for Sivers effect in \boldmath $A_{UL}^{\sin\phi}$
	asymmetries?}
\label{Sec-4:Sivers}

The introduction of a Gaussian distribution of transverse momentum
allows us to consider in the analysis the contribution of the Sivers effect.
In the previous Sections we set the Sivers function to zero based on the
results from the $\chi$QSM. However, despite the successes of this model,
we have to admit that so far it has not been used and tested in the arena
of unintegrated parton distributions.
The vanishing of the Sivers function and other T-odd distributions in the
$\chi$QSM and a large class of other chiral models demonstrates the
limitations of such models \cite{Pobylitsa:2002fr}.
From the point of view of the $\chi$QSM and the instanton model of the QCD
vacuum T-odd distribution functions appear to be suppressed with respect to
the T-even ones \cite{Efremov:2003tf}.
The instanton vacuum suppression mechanism was demonstrated to be strong in
the case of $\widetilde{g}_T^a(x)$ \cite{Balla:1997hf} which recently was
confirmed experimentally \cite{Anthony:2002hy}.
In the case of the Sivers function this mechanism has not yet been studied
rigorously but concluded on general grounds and could be less pronounced
\cite{Efremov:2003tf}.
It is therefore instructive to investigate whether anything can be
concluded on the magnitude of the Sivers effect from the HERMES data
on $A_{UL}^{\sin\phi}$.

At first glance Figs.~2a-2f may give the impression that the Collins effect
alone is able to explain nicely the data -- leaving no or little room for the
Sivers effect and implying that the Sivers distribution function is small.
However, we shall see in the following that this needs not be the case.
For that let us concentrate on the asymmetries that are least
sensitive to the assumption of favoured flavour fragmentation, namely
$A_{UL}^{\sin\phi}$ for $\pi^+$ and $\pi^0$ from the proton target.
Under the assumption of favoured flavour fragmentation,
one can expect the total $A_{UL}^{\sin\phi}(z)$ to behave as
\be\label{Eq:AUL-z-final-with-Sivers}
	A_{UL}^{\sin\phi}(z,\,\pi) =
	C_L(\pi,\,\,{\rm target}) \; a_{\rm Gauss}^{(1/2)}(z) 
	+ 
	B_{\rm
	Siv}(\pi)\Big/{\sqrt{1+\la{\bf K}_T^2(z)\ra/(z^2 \la{\bf p}_T^2\ra})}
	\;.\ee
where $B_{\rm Siv}$ is a constant independent of $z$ due to the Sivers effect.

%--- FIGURE 4: AUL with Sivers (pi+,pi0;proton) ---------------------------
	\begin{figure}[t!]
\begin{tabular}{cc}
	\includegraphics[width=5.2cm,height=5cm]{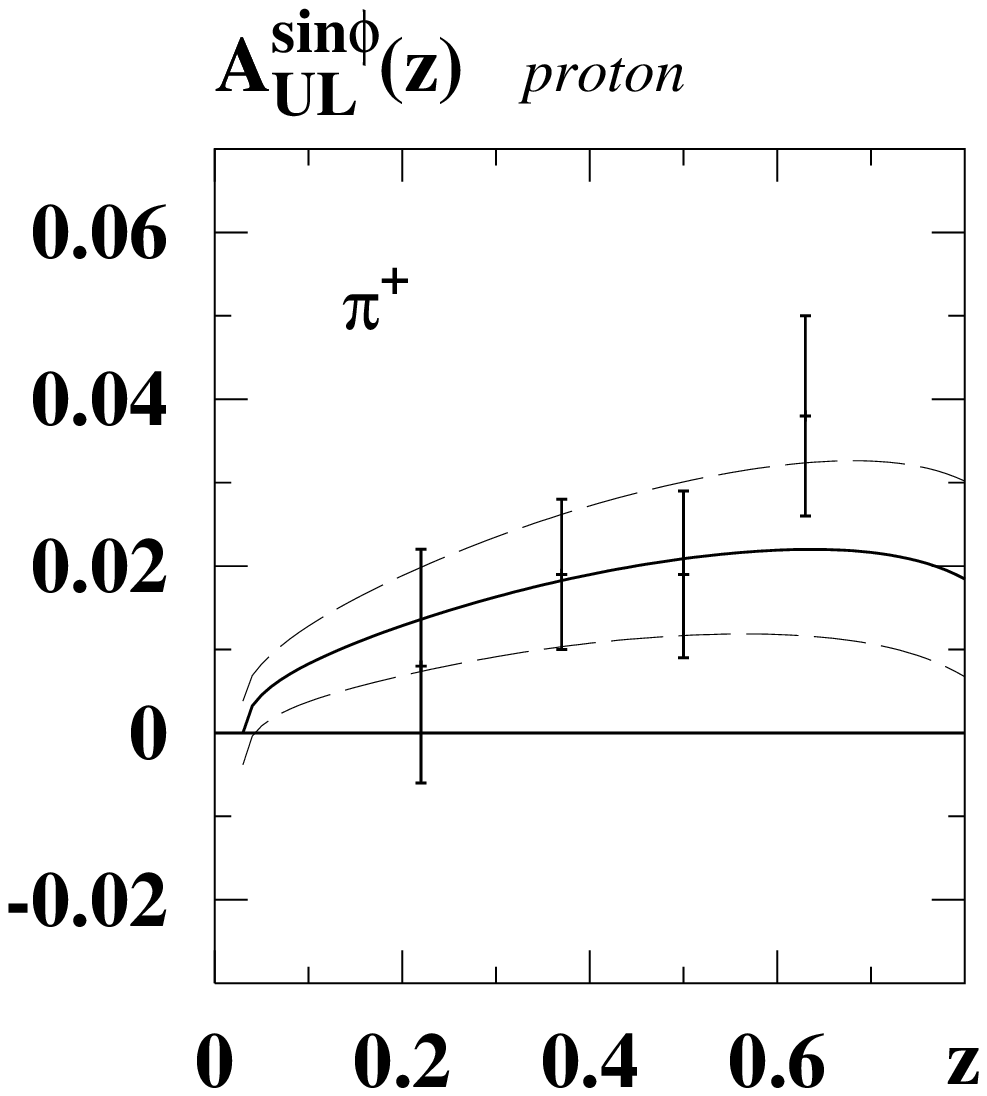} &
	\includegraphics[width=5.2cm,height=5cm]{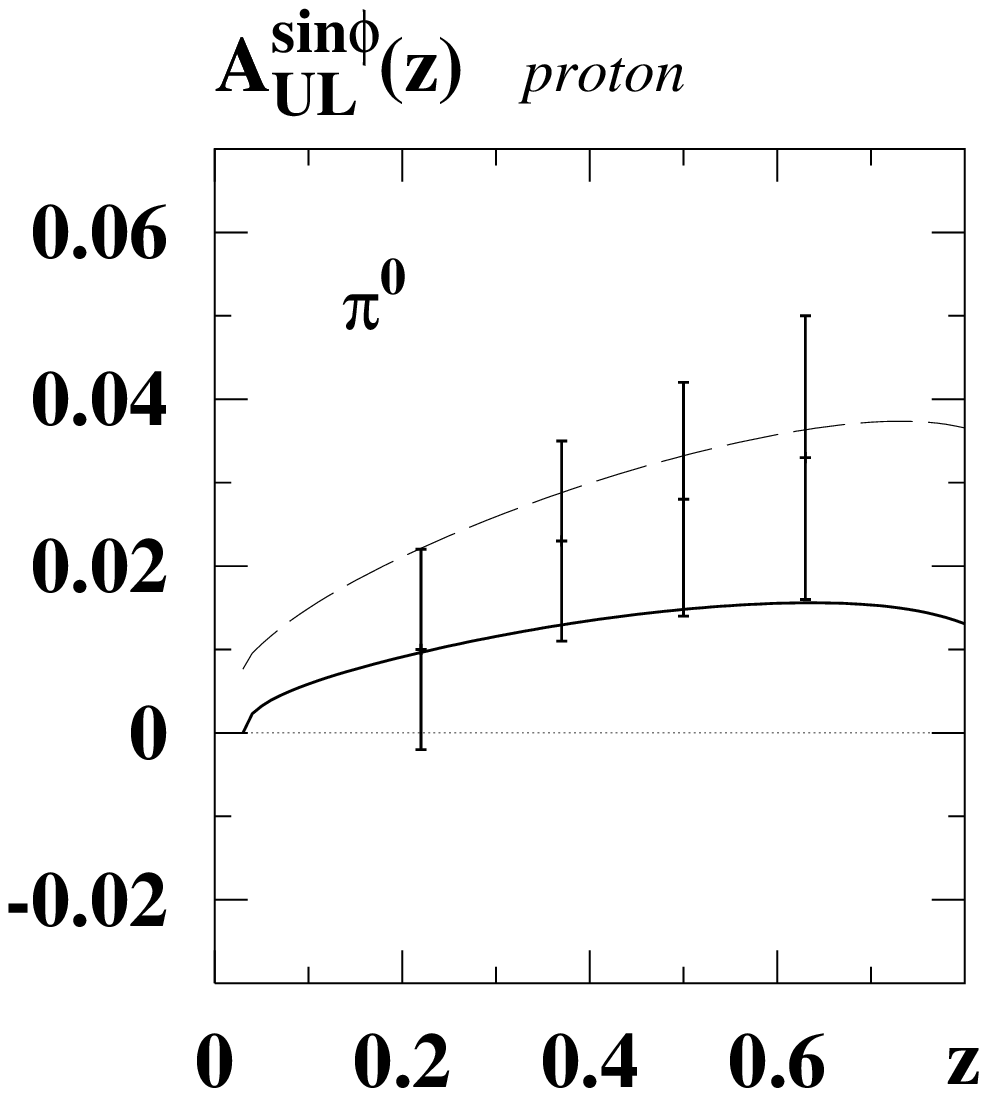}
	\cr
	{\bf a} &
	{\bf b}
\end{tabular}
	\caption{\footnotesize\sl
	Rough estimate of the possible magnitude of the Sivers effect in
	the azimuthal single spin asymmetries $A_{UL}^{\sin\phi}(z)$ from
	a proton target for ({\bf a}) $\pi^+$ and ({\bf b}) $\pi^0$.
	The thin solid lines are the central values shown in Figs.~2a and 2b.
	The dashed lines are the shifts of the central values due to the
	Sivers effect according to
	Eqs.~(\ref{Eq:AUL-z-final-with-Sivers},~\ref{Eq:B-Siv-range}).}
	\end{figure}
%--- END FIGURE 4. --------------------------------------------------------

Figs.~4a and 4b show that reasonable descriptions of $A_{UL}^{\sin\phi}(z)$
for $\pi^+$ and $\pi^0$ is possible with roughly
\be\label{Eq:B-Siv-range}
	-\,\frac{1}{80} \lsim B_{\rm Siv}(\pi^+) \lsim \frac{1}{80} \;,\;\;\;
	0 \lsim B_{\rm Siv}(\pi^0) \lsim \frac{1}{40}
\ee
The different ranges of $B_{\rm Siv}$ for $\pi^+$ and $\pi^0$
could reflect the flavour dependence of the Sivers function.

From the expression (\ref{Eq:sigmaUT-02}) we obtain roughly
but with sufficient accuracy for our purposes
\ba
	B_{\rm Siv}=\frac{2\,\int\di x\,\di y\;\sin\Theta_S\,(1-y+y^2/2)Q^{-4}
        \sum_a^\pi e_a^2\, x \,f_{1T}^{\perp (1/2)a}(x)}
	{\int\di x\,\di y\;(1-y+y^2/2)Q^{-4}\sum_b^\pi e_b^2 \,x\, f_1^b(x)}
	\approx
	\frac{2\,\la\sin\Theta_S\ra
	      \int\di x\,\sum_a^\pi e_a^2\, x \,f_{1T}^{\perp (1/2)a}(x)}
	     {\int\di x\sum_b^\pi e_b^2\, x \,f_1^b(x)} \,.\;
\ea
We estimate $\la\sin\Theta_S\ra \approx$
$\sqrt{2 M_N^2\la x\ra(1-\la y\ra)/(s \la y\ra)} = 0.05$
(with $\la x\ra = 0.09$ and $\la y\ra = 0.57$
\cite{Airapetian:1999tv,Airapetian:2001eg}).
This gives
\ba
	\frac{\int\di x\sum_a^\pi e_a^2 x f_{1T}^{\perp (1/2) a}(x)}
	     {\int\di x\sum_b^\pi e_b^2 x f_1^b(x)}
	\approx 40 \,B_{\rm Siv}\, ,
\ea
and we obtain the bounds
\ba\label{Sivers-bounds}
	-\,\frac12 \lsim
	\frac{\int\di x\sum_a^{\pi^+}e_a^2\,x\,f_{1T}^{\perp (1/2)a}(x)}
	     {\int\di x\sum_b^{\pi^+}e_b^2\,x\,f_1^b(x)} \lsim \frac12
	\;\;,\;\;\;\;\;
	 0 \lsim
	\frac{\int\di x\sum_a^{\pi^0}e_a^2\,x\,f_{1T}^{\perp (1/2)a}(x)}
	     {\int\di x\sum_b^{\pi^0}e_b^2\,x\,f_1^b(x)} 
	\lsim 1 \;\;.
\ea
However, the Sivers function obeys the positivity bound
$|f_{1T}^{\perp (1/2)a}(x)|\le 1/2 f_1^a(x)$ \cite{Bacchetta:1999kz},
such that for any pion
\be\label{Sivers-positivity-bound}
	-\,\frac12 \le
	\frac{\int\di x\sum_a^{\pi}e_a^2\,x\,f_{1T}^{\perp (1/2)a}(x)}
	     {\int\di x\sum_b^{\pi}e_b^2\,x\,f_1^b(x)} \le \frac12 \,. \ee
Thus, the bounds (\ref{Sivers-bounds}) do not provide any useful information
on the Sivers function except for the lower
bound in the $\pi^0$ case.
Our procedure to obtain the ``bounds'' (\ref{Sivers-bounds}) is admittedly
crude and model dependent. E.g., by assuming a somehow lower value for
$\la{\bf p}_T^2\ra$ we could have obtained a negative lower 
bound in the $\pi^0$ case in Eq.~(\ref{Sivers-bounds}). 
Still it allows to learn an important lesson.
The HERMES data on $A_{UL}^{\sin\phi}$ can be well described without the
Sivers effect by invoking the Collins effect alone. However, from this
observation we by no means can conclude that the Sivers effect is small.
Indeed, a Sivers effect as large as allowed by the positivity bound
\cite{Bacchetta:1999kz}  -- in particular as large as required
to explain \cite{Anselmino:1994tv,D'Alesio:2003dv} the large SSA
in $p^\uparrow p\to \pi X$ \cite{Adams:1991cs} --
could {\sl not} be resolved at present within the statistical error bars
of the HERMES data \cite{Airapetian:1999tv,Airapetian:2001eg}.

%====== SECTION 6: PREDICTION =====================================
\section{Predictions for the \boldmath $A_{UT}^{\sin(\phi+\phi_S)}$
         Collins asymmetry}
\label{Sec-5:AUT}

In the previous Sections we have seen that the HERMES data on the
$A_{UL}^{\sin\phi}(z)$ asymmetries can well be described in terms of the
Collins effect, however, they are compatible with a sizeable Sivers effect,
too.
Azimuthal single spin asymmetries from a transversely polarized target are key
observables, since they allow to cleanly distinguish the Collins and Sivers effect
by the different azimuthal distribution of the produced pions, schematically
\be
	A_{UT} \propto
	\left(\frac{\di\sigma^\uparrow  }{S^\uparrow  }
	     -\frac{\di\sigma^\downarrow}{S^\downarrow}\right)
	\propto  \sin(\phi+\phi_S) \cdot (\mbox{Collins effect})
		+\sin(\phi-\phi_S) \cdot (\mbox{Sivers effect})
\ee
where $^{\uparrow(\downarrow)}$ denote the transverse with respect to the
lepton beam target polarizations and $\phi_S$ is the azimuthal angle of the
target polarization vector, see Fig.~1. Thus, by considering appropriate
weights \cite{Boer:1997nt} both effects can be separated.
(In the longitudinal target polarization experiments $\phi_S$ was $0$
or $\pi$ and dropped out from the weighting factor $\sin\phi$.)

Transverse target polarization experiments are in progress at HERMES
\cite{Makins:uq} and COMPASS \cite{LeGoff:qn}.
In this Section we shall estimate the transverse target single spin
asymmetry due to the Collins effect for the HERMES experiment.
Defining $A_{UT}^{\sin(\phi+\phi_S)}$ in analogy to Eq.~(\ref{Eq:AUL-00}) and
using the same assumptions as in Section~\ref{Sec-3.1:neglect-pT} we obtain
\be\label{Eq:AUT-z}
	A_{UT}^{\sin(\phi+\phi_S)}(z,\,\pi,\,{\rm target}) =
	C_T(\pi,\,\,{\rm target}) \; a^{(1/2)}(z) \;
\ee
with $a^{(1/2)}(z)$ as defined in Eq.~(\ref{Eq:universal-a^(1/2)(z)}) while if
we adopt a Gaussian ansatz as done in Section~\ref{Sec-3.2:AUL-Gauss} we obtain
\be\label{Eq:AUT-z-gauss}
	A_{UT}^{\sin(\phi+\phi_S)}(z,\,\pi,\,{\rm target}) =
	C_T(\pi,\,\,{\rm target}) \; a^{(1/2)}_{\rm Gauss}(z) \;,
\ee
with $a^{(1/2)}_{\rm Gauss}(z)$ as defined in
Eq.~(\ref{App:Gaussian-universal-a(z)}). The constant $C_T$ turns out to be
\be\label{Eq:CT-pi-target}
	C_T(\pi,\,{\rm target}) = 2\;
	\frac{\int\di x\di y\cos\Theta_S(1-y)/Q^4
	      \sum_a^\pi e_a^2\,xh_1^{a/{\rm target}}(x)}
	     {\int\di x\di y\,(1-y+y^2/2)/Q^4
	      \sum_a^\pi e_a^2\,xf_1^{a/{\rm target}}(x)} \;.\ee
Taking the cuts as in the longitudinal target polarization experiment,
Eq.~(\ref{Eq:exp-cuts}), we obtain for $C_T(\pi,\,\,{\rm target})$
the results quoted in Table~1.
Figs.~5a-5f show the results for $A_{UT}^{\sin(\phi+\phi_S)}(z,\,\pi)$
for the proton and deuteron target, for the two different assumptions on the
transverse momentum distribution: no intrinsic $p_T$ (dotted line) and a
Gaussian $p_T$ (solid line).

%--- FIGURE 5: AUT proton+deuteron -----------------------------
	\begin{figure}[t!]
\begin{tabular}{ccc}
	\includegraphics[width=5.2cm,height=5cm]{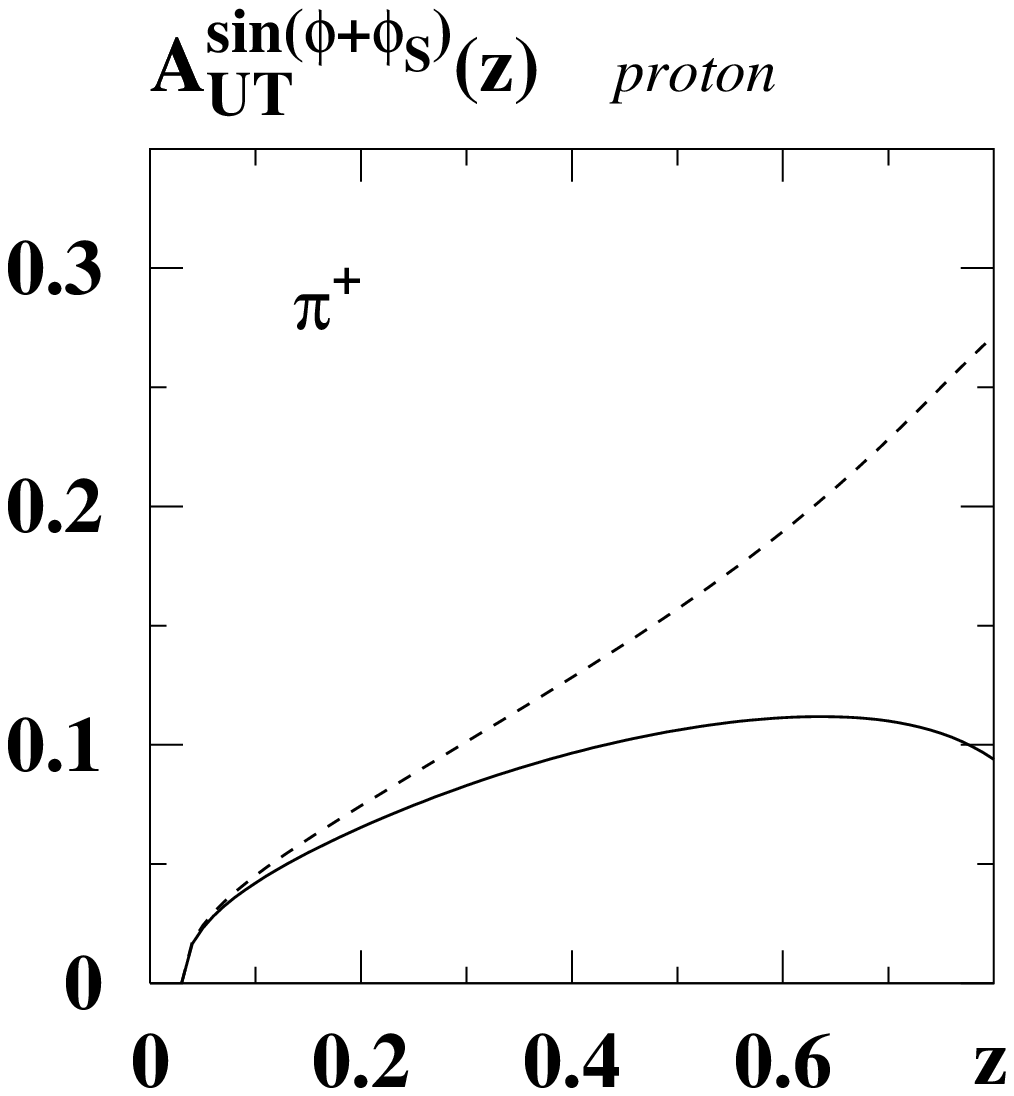} &
	\includegraphics[width=5.2cm,height=5cm]{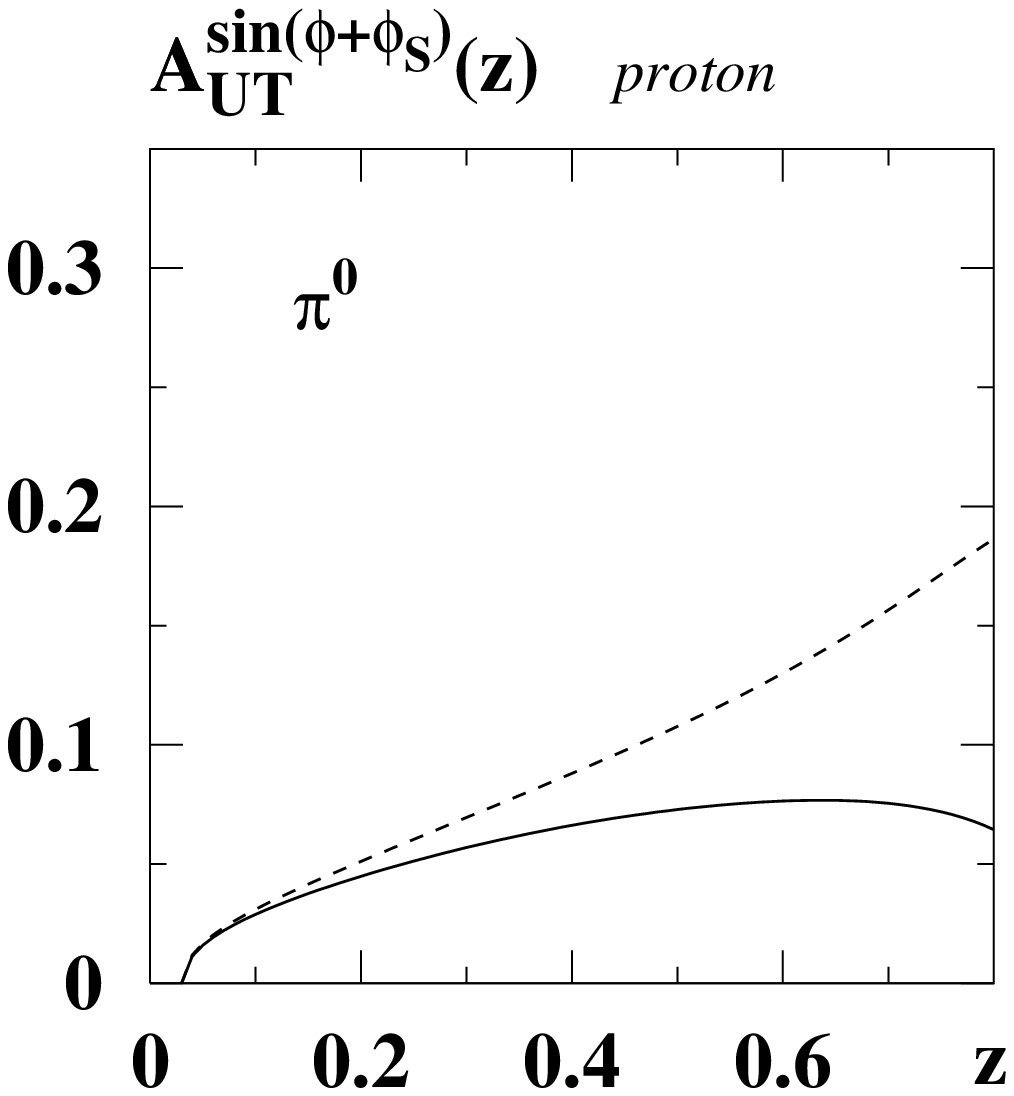} &
	\includegraphics[width=5.2cm,height=5cm]{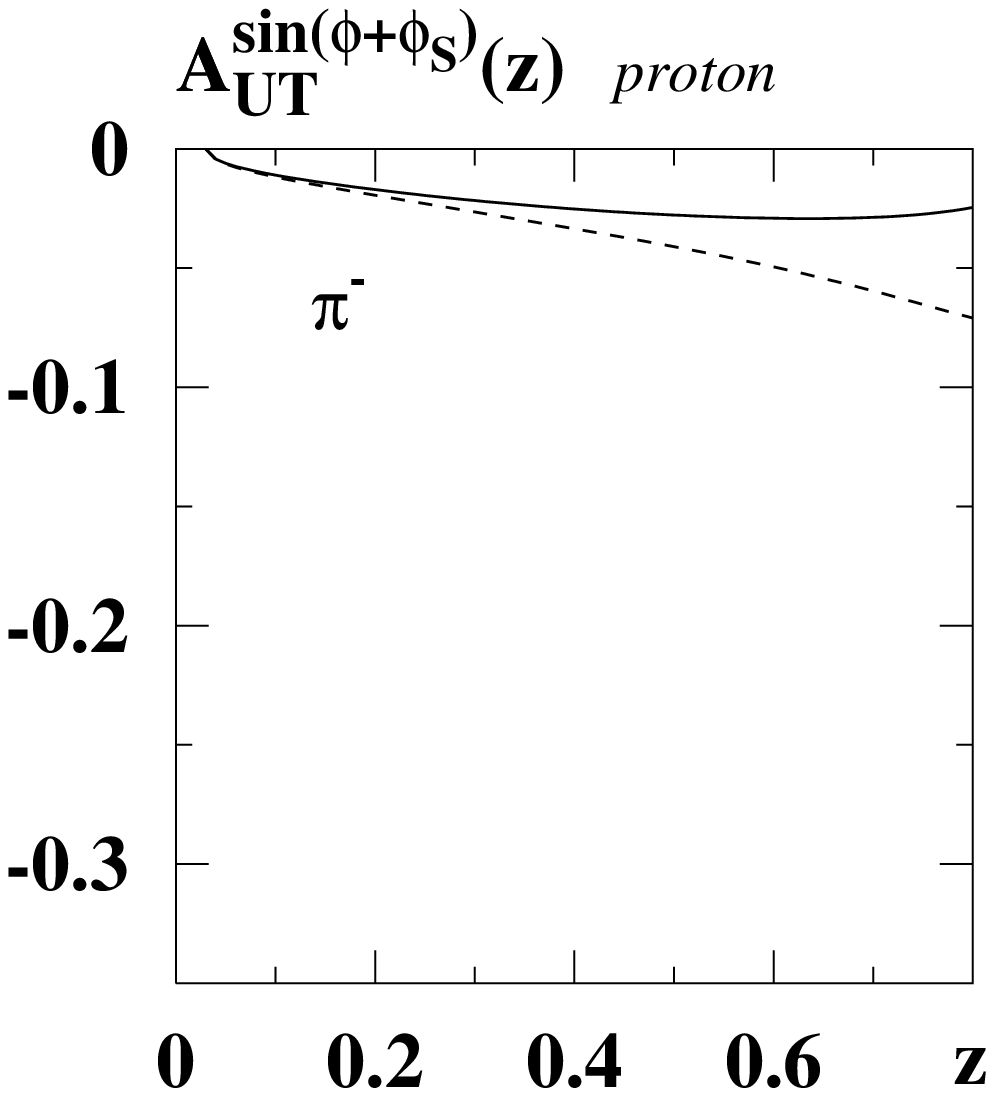}
	\cr
	{\bf a} &
	{\bf b} &
	{\bf c} \cr
	\includegraphics[width=5.2cm,height=5cm]{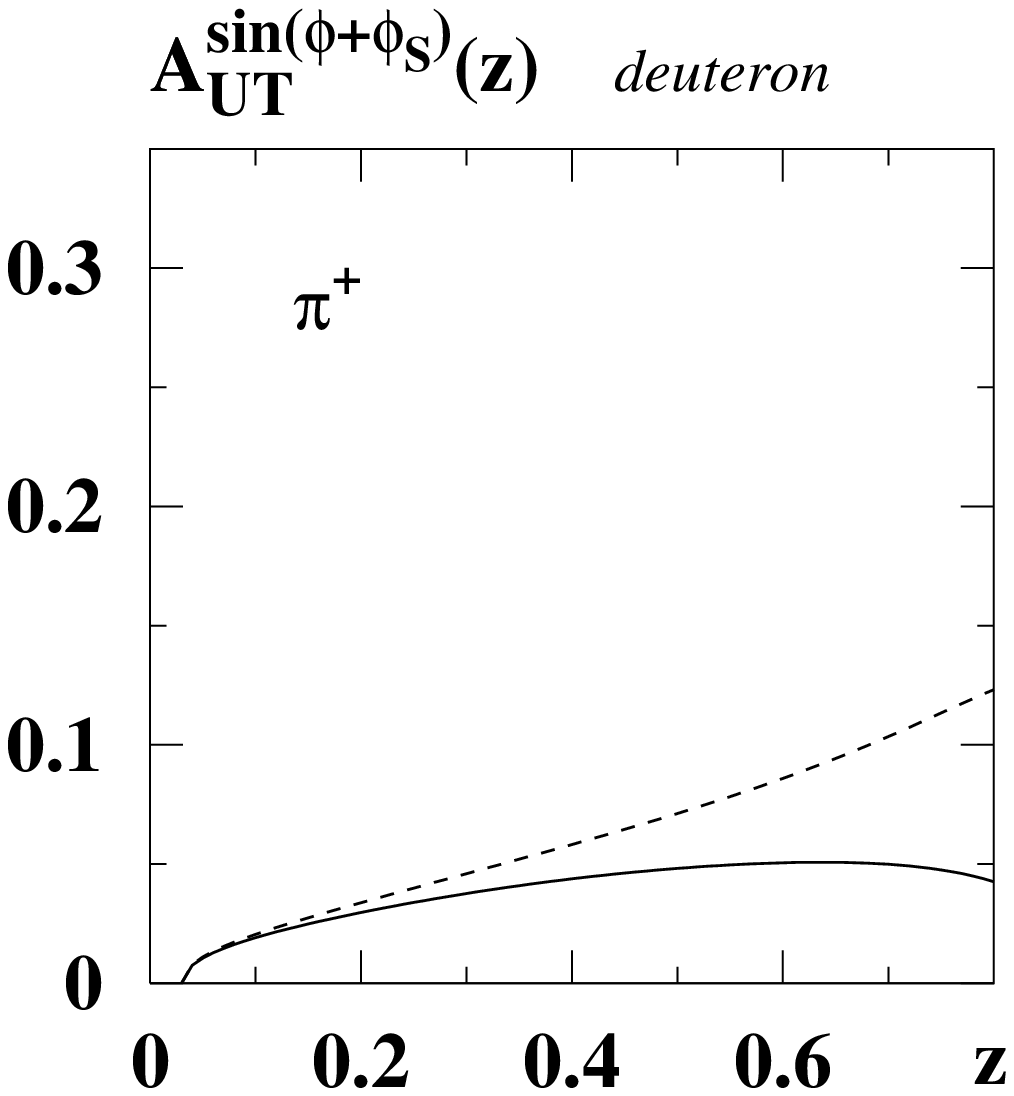} &
	\includegraphics[width=5.2cm,height=5cm]{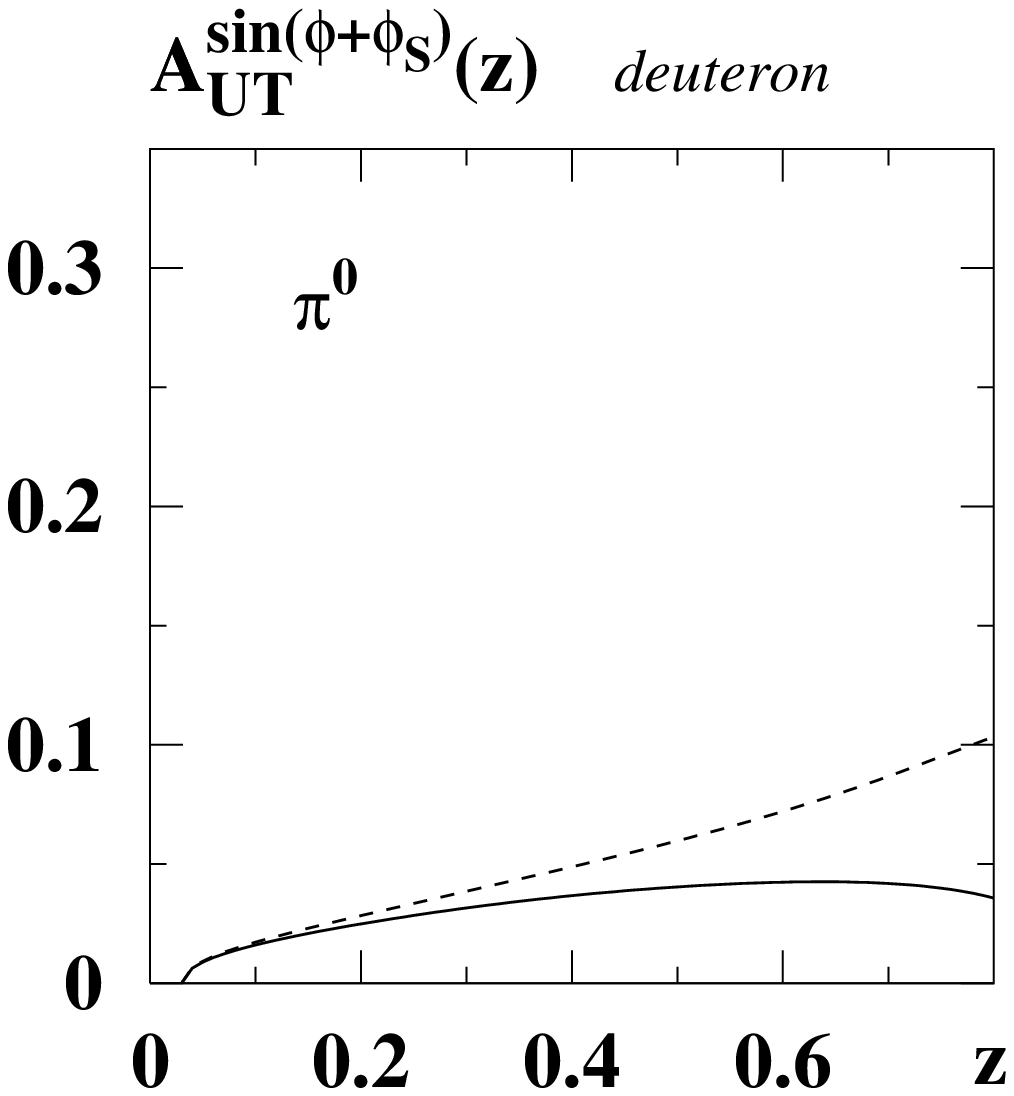} &
	\includegraphics[width=5.2cm,height=5cm]{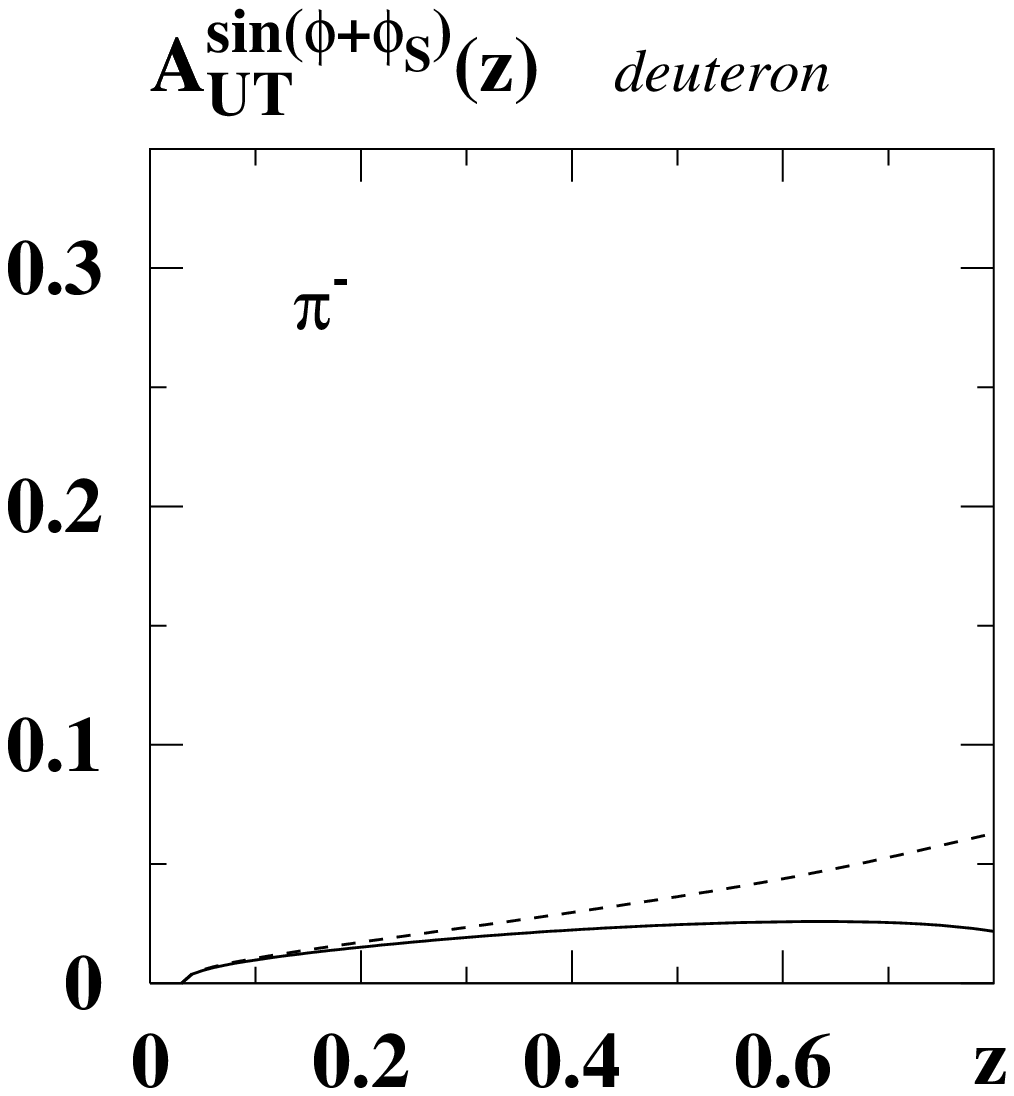}
	\cr
	{\bf d} &
	{\bf e} &
	{\bf f}
\end{tabular}
	\caption{\footnotesize\sl
	The asymmetries $A_{UT}^{\sin(\phi+\phi_S)}(z)$ for pion
	production from a transversely polarized proton target at HERMES
	for the assumptions that intrinsic transverse parton momenta in the
	target are follow a Gaussian distribution (solid lines) and are
	negligible (dashed lines).}
	\end{figure}
%--- END FIGURE 5. --------------------------------------------------------

Finally, we consider the weighted
asymmetry $A_{UT}^{\sin(\phi+\phi_S)P_{\pi\perp}/m_\pi}$ \cite{Boer:1997nt}.
Assuming favoured flavour and isospin invariance relations among the
fragmentation functions, but without any assumptions on the transverse
momentum distribution, the asymmetry takes the form
\be\label{Eq:AUT-z-corr-weight}
	A_{UT}^{\sin(\phi+\phi_S)P_{\pi\perp}/m_\pi}(z,\,\pi,\,{\rm target}) =
	C_T(\pi,\,\,{\rm target}) \; a^{(1)}(z) \;
\ee
with the same constant $C_T$ as in $A_{UT}^{\sin(\phi+\phi_S)}$, cf.\
Eq.~(\ref{Eq:CT-pi-target}), but a different ``universal function''
$a^{(1)}(z)$ defined as
\be\label{X}
	a^{(1)}(z) = \frac{z H_1^{\perp (1)}(z)}{D_1(z)} \;.
\ee
where
\begin{equation}
 H_1^{\perp(1)}(z) \equiv
\int \di^2 {\bf K}_{T}\, \frac{{\bf K}_{T}^2}{2 z^2 m_{\pi}^2}\,
H_1^{\perp}(z,{\bf K}_{T}^2) \,.
\end{equation}
Figs.~6a and 6b show the predictions of our model for the weighted
asymmetry for the proton and deuteron target, respectively.

%--- FIGURE 6: AUT proton --------------------------------------
	\begin{figure}[t!]
\begin{tabular}{cc}
	\includegraphics[width=5.2cm,height=5cm]{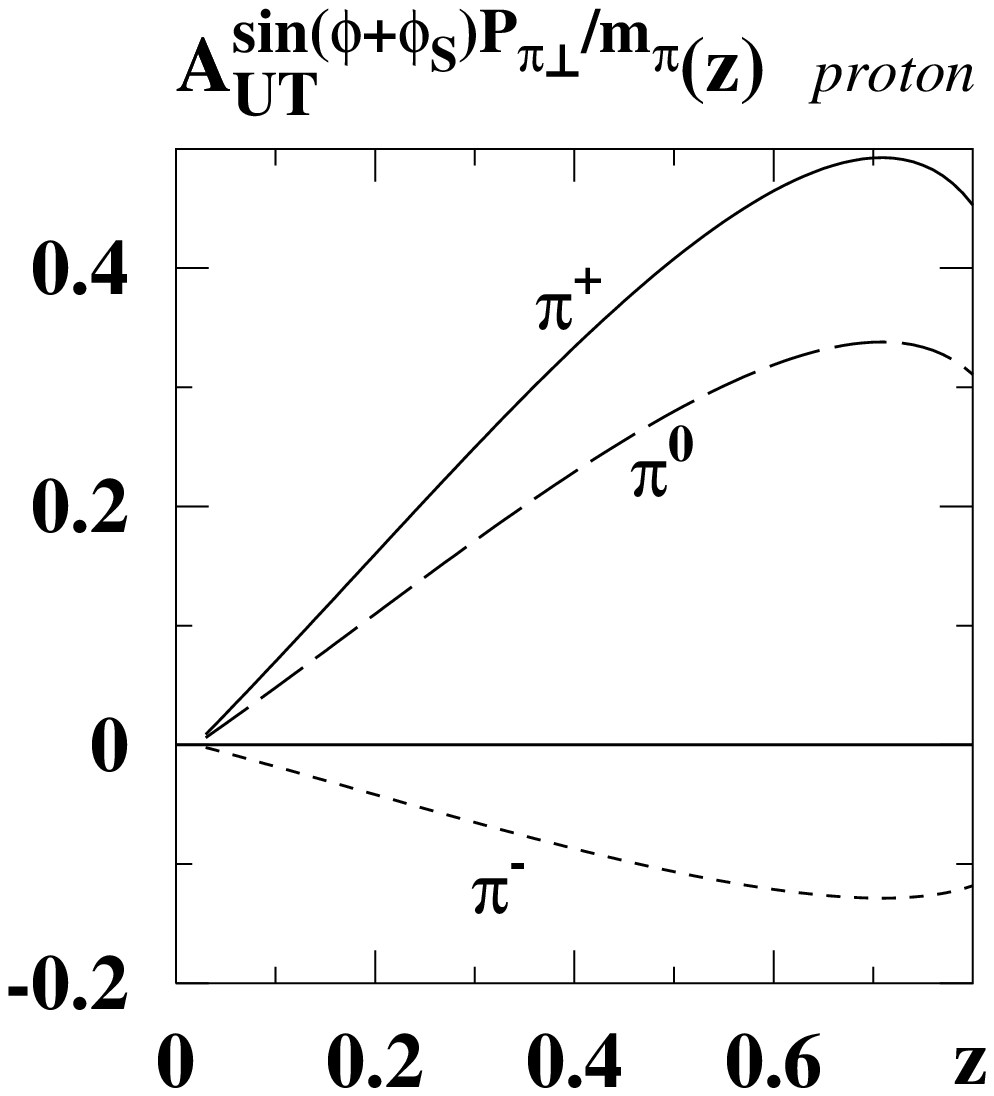} &
	\includegraphics[width=5.2cm,height=5cm]{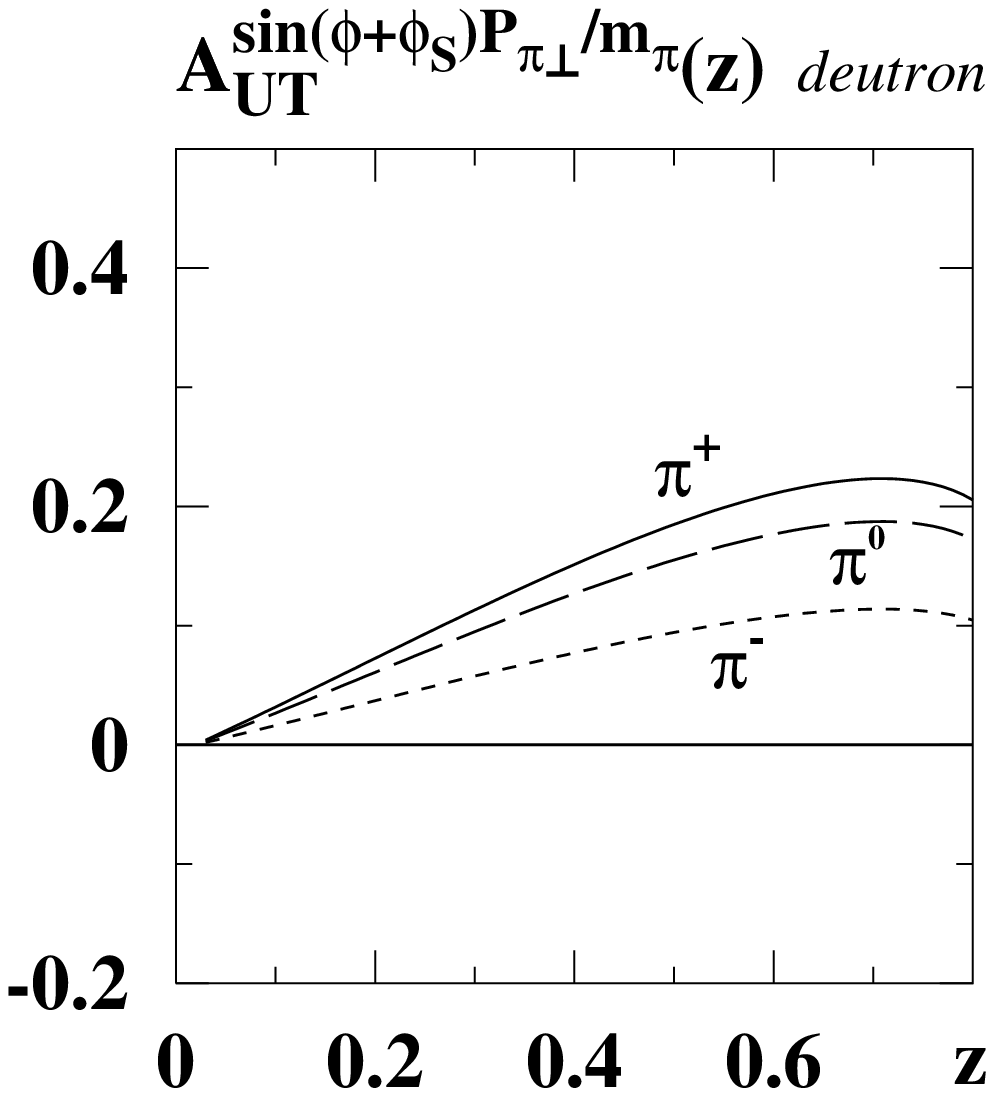} \cr
	{\bf a} &
	{\bf b}
\end{tabular}
	\caption{\footnotesize\sl
	The asymmetries transverse momentum weighted asymmetries
	$A_{UT}^{\sin(\phi+\phi_S)P_{\pi\perp}/m_\pi}(z)$ as functions of $z$
	for pion production from a transversely polarized proton (a) and
	deuteron (b) target for kinematics of the HERMES experiment.}
	\end{figure}
%--- END FIGURE 6. --------------------------------------------------------

%====== SECTION 6: CONCLUSIONS =====================================
\section{Comment on the $x$-dependence of \boldmath
	$A_{UL}^{\sin\phi}$ and $A_{UT}^{\sin(\phi+\phi_S)}$}
\label{Sec:AUL-AUT-x}

The $x$-dependence of the asymmetries $A_{UL}^{\sin\phi}(x)$ and
$A_{UT}^{\sin(\phi+\phi_S)}(x)$ is determined by the predictions for the
chirally odd distribution functions from the chiral quark-soliton and
instanton vacuum model \cite{h1-model,Dressler:1999hc},
while their overall normalization is fixed by the predictions for
the Collins fragmentation function from the calculation in the
Georgi-Manohar model \cite{Bacchetta:2002tk}.
Also in Refs.~\cite{Efremov:2001cz,Efremov:2001ia,Efremov:2003eq} the chiral
quark-soliton and instanton vacuum model predictions were explored to study the
$x$-dependence of the HERMES data using, however, different information on the
Collins fragmentation function. Let us discuss how much the results presented
in Refs.~\cite{Efremov:2001cz,Efremov:2001ia,Efremov:2002ut,Efremov:2003eq} 
would be altered, if we used instead the Georgi-Manohar model results for 
the Collins fragmentation function \cite{Bacchetta:2002tk}.

In Refs.~\cite{Efremov:2001cz,Efremov:2001ia,Efremov:2002ut,Efremov:2003eq}
a different treatment of transverse parton momenta was employed\footnote{
	In Refs.~\cite{Efremov:2003eq} a Gaussian distribution of parton
	transverse momenta in the target was assumed with a width
	$\la|{\bf p}_T|\ra=0.4\,{\rm GeV}$ from
	\cite{Martin:1998sq,Abreu:1996na}. However,
	Eq.~(\ref{Eq:Phperp-in-SIDIS}) was not used to relate the involved
	transverse momenta, but $z{\bf K}_T$ was directly identified with
	the transverse momentum ${\bf P}_{h\perp}$ of the produced hadron.}
and a different normalization of the Collins fragmentation function was used.
In Refs.~\cite{Efremov:2001cz,Efremov:2001ia,Efremov:2002ut,Efremov:2003eq}
the quantity
\be\label{Eq:compare-I}
	\Biggl[\frac{1}{2\la z\ra
	\sqrt{1+\la{\bf p}_T^2\ra/\la{\bf K}_T^2\ra}}\,
	\frac{\la H_1^\perp\ra}{\la D_1\ra}\Biggr]_{
	\mbox{\footnotesize
	\cite{Efremov:2001cz,Efremov:2001ia,Efremov:2002ut,Efremov:2003eq}}}
	= \cases{
	0.12& with $|\la H_1^\perp\ra|$   from DELPHI \cite{Efremov:1998vd},\cr
	0.14& with $ \la H_1^\perp\ra $ ``from HERMES'' \cite{Efremov:2001cz}}
\ee
corresponds to the following $z$-averaged quantities in our approach
(the integration goes over $0.2\le z\le 0.7$)
\ba
	\la a^{1/2}(z)\ra &\equiv&
	\frac{\int\di z\,H_1^{\perp (1/2)}(z)}{\int\di z\,D_1(z)} = 0.140 \; ,
	\nonumber\\
	\la a^{1/2}_{\rm Gauss}(z)\ra &\equiv&
	\frac{\int\di z\,H_1^{\perp (1/2)}(z)/
	      \sqrt{1+z^2 \la{\bf p}_T^2\ra/\la{\bf K}_T^2(z)\ra}}
	      {\int\di z\,D_1(z)} = 0.104 \; .
	\label{Eq:compare-II}
\ea
The first number in (\ref{Eq:compare-I}) follows from using the value
from the DELPHI experiment on $e^+e^-$ annihilation at the $Z^0$ peak
\cite{Efremov:1998vd}, which provided the first experimental indication
for $H_1^\perp$. The second number follows from the value extracted in
Ref.~\cite{Efremov:2001cz} from the HERMES (SIDIS) data
\cite{Airapetian:1999tv,Airapetian:2001eg} assuming the chiral quark-soliton
model results for the unknown distribution functions.
These numbers are numerically close to each other {\sl and} to the values in
(\ref{Eq:compare-II}) from the model calculation of \cite{Bacchetta:2002tk}.
This means that the approach considered in this work describes the
$x$-dependence of the azimuthal single spin asymmetries measured at HERMES
with a longitudinally polarized target similarly well as the approach of
Refs.~\cite{Efremov:2001cz,Efremov:2001ia}, and predicts numerically similar
effects for CLAS longitudinal and HERMES transverse target polarization
experiments \cite{Efremov:2002ut,Efremov:2003eq}.

In this context it should be mentioned  that using the DELPHI ($e^+e^-$
annihilation) result \cite{Efremov:2001cz} in the description of the HERMES
(SIDIS) experiment has -- apart from disregarding possible Sudakov suppression
effects \cite{Boer:2001he} -- the drawback of presuming universality of the
Collins fragmentation function which has recently been questioned
\cite{Boer:2003cm}.
(Though, no indications for process dependence have been observed in leading
order perturbation theory \cite{Metz:iz}. Also in the model calculations
\cite{Bacchetta:2002tk,Gamberg:2003eg,Bacchetta:2003xn} the Collins function
is universal.)
It is remarkable that the numbers in (\ref{Eq:compare-I},~\ref{Eq:compare-II})
from different experiments and model calculations referring to different
scales are numerically close to each other.

%====== SECTION 6: CONCLUSIONS =====================================
\section{Summary and conclusions}
\label{Sec-6:conclusions}

One aim of this work was to study how much of the HERMES data on azimuthal
single spin asymmetries $A_{UL}^{\sin\phi}$ from a longitudinally polarized
target can be explained by chiral physics, relying on effective chiral
quark and Goldstone boson degrees of freedom. The only T-odd effect
which is needed to explain single spin asymmetries and can be modelled
in terms of the chiral degrees of freedom is the Collins effect.
The Sivers effect requires to take into account explicitly
gluonic degrees of freedom\footnote{
	This statement is based on the observation that one cannot
	{\sl simply} model T-odd distribution functions directly in
	chiral models \cite{Pobylitsa:2002fr}. However, it cannot be
	excluded that it is possible to find an effective representation
	of the gauge link in terms of chiral degrees of freedom on the basis
	of a formalism which is able to relate gluonic degrees of freedom
	to effective quark degrees of freedom -- such as the instanton
	formalism developed in Ref.~\cite{Diakonov:1996qy}.}
as, e.g., it is done in  models with one-gluon exchange
\cite{Brodsky:2002pr,Gamberg:2003ey,Yuan:2003wk,Bacchetta:2003rz}.

In particular, we focused on the $z$-dependence of the asymmetries
$A_{UL}^{\sin\phi}$ since the Collins and Sivers effect have different
dependence on $z$ and could -- at least in principle\footnote{
	It will be interesting to see whether this possibility can
	be practically explored in the CLAS and Hall-A experiments at
	Jefferson Lab -- where the high luminosity promises a large
	statistics and high precision data.}
 -- be distinguished in this way.
For the Collins fragmentation function we took the prediction from
the calculation in the Georgi-Manohar model \cite{Bacchetta:2002tk}.
The required information on the integrated chirally odd distribution
functions, which provides the overall normalization, we took from the chiral
quark-soliton and instanton vacuum model \cite{h1-model,Dressler:1999hc}.
We used two different assumptions on intrinsic transverse parton momenta in
the target, namely that they vanish or are Gaussian distributed, respectively.
Both assumptions yield a reasonable description of the HERMES data
\cite{Airapetian:1999tv,Airapetian:2001eg,Airapetian:2002mf}.

The good description of the HERMES data in terms of the Collins
effect only, as observed also previously in other analyses
\cite{DeSanctis:2000fh,Anselmino:2000mb,Efremov:2000za,Ma:2002ns,Efremov:2001cz,Efremov:2001ia},
could give rise to the suspicion that the Sivers effect is small.
However, by making a crude qualitative estimate, we have shown that within the
present experimental error bars even a sizeable Sivers effect could be hidden
-- as observed in \cite{Efremov:2003tf} on the basis of independent arguments.

Whether or not chiral degrees of freedom -- as considered in our models --
are able to model realistically the Collins effect can cleanly be tested
in the transverse target polarization experiments which are in progress
at HERMES \cite{Makins:uq} and COMPASS \cite{LeGoff:qn}.
For that we have predicted the asymmetry $A_{UT}^{\sin(\phi+\phi_S)}(z)$ for
the HERMES kinematics which appears to be large, of order $20\%$ for $\pi^+$
from the proton target (to be compared with $A_{UL}^{\sin\phi}\sim(2-4)\%$).
In the kinematics of the COMPASS experiment the effect is similarly large.

A measurement of $A_{UT}^{\sin(\phi+\phi_S)}$ asymmetries in the HERMES and
COMPASS experiments of comparable magnitude to what predicted here would
suggest that chiral symmetry breaking -- in the way it is considered in our
approach -- can be used as a guideline to estimate the Collins effect, 
thus offering a valuable contribution to the understanding of
the theory and phenomenology of single spin asymmetries.

\paragraph{Comment added.}
After this work was basically completed the first results from the HERMES
transverse polarization proton target experiment have been released
\cite{HERMES-AUT-prelim}. The preliminary data for $\pi^+$ are compatible 
with our predictions. (Note that the definition of $A_{UT}$ used here 
includes an extra factor of 2 compared to the definition used by the 
HERMES collaboration \cite{HERMES-AUT-prelim}.)
The asymmetries for $\pi^0$ and more clearly $\pi^-$ seem not to be 
compatible with our estimates. This probably indicates that taking only 
favoured fragmentation into account is not sufficient to describe the 
Collins effect.

\paragraph{Acknowledgements.}
We would like to thank A.~V.~Efremov, K.~Goeke, A.~Metz and M.~V.~Polyakov
for fruitful discussions and B.~Dressler for providing the evolution code.
This work has partly been performed under the contract
HPRN-CT-2000-00130 of the European Commission. The work of A.~B. has been
partially supported by the BMBF.

%====== APPENDIX ===================================================
   \appendix
   \setcounter{equation}{0}
   \def\theequation{A.\arabic{equation}}

%===================  REFERENCES =====================================

\end{document}